\documentclass[oneside,english]{elsarticle}

\usepackage{amsmath}

\usepackage[T1]{fontenc}
\usepackage[latin9]{inputenc}
\usepackage{geometry}
\usepackage{float}
\usepackage{textcomp}
\usepackage{amsthm}
\usepackage{amstext}
\usepackage{amssymb}
\usepackage{graphicx}
\usepackage{esint}
\usepackage{subscript}

\makeatletter

\providecommand{\tabularnewline}{\\}

\everymath{\displaystyle}

\usepackage{chngcntr}
\counterwithout{figure}{section}

\makeatother

\usepackage{babel}

\begin{document}

\begin{frontmatter}

\title{Expansion of a spherical dust gas\\
-- the cosmological conundrum}


\author[rvt]{Ingo Müller\corref{cor1}}
\ead{ingo.mueller@alumni.tu-berlin.de}

\author[rvt]{Wolf Weiss}
\ead{wolf.weiss@alumni.tu-berlin.de}

\cortext[cor1]{Corresponding author}

\address[rvt]{TU Berlin,Straße des 17. Juni 135, 10623 Berlin, Germany}

\author{}

\address{}

\begin{abstract}
The universe is viewed as a dust gas filling a sphere and floating
in infinite empty space. Einstein\'{ }s gravitational equations are
applied to this case together with appropriate boundary values. The
equations are solved for initial conditions chosen so as to describe
the observed Hubble diagram. We find that the solution is not unique
so that more astronomical observations are needed. However, those
solutions which \textit{were} found do not exhibit an accelerated
expansion of the universe, nor -- obviously then -- do they need the
notion of a dark energy driving such an expansion.

We present this study as an alternative to the prevailing Robertson-Walker
cosmology.
\end{abstract}

\end{frontmatter}

\section{Introduction}

\subsection*{The FRW model}

\begin{figure}[h]
\includegraphics{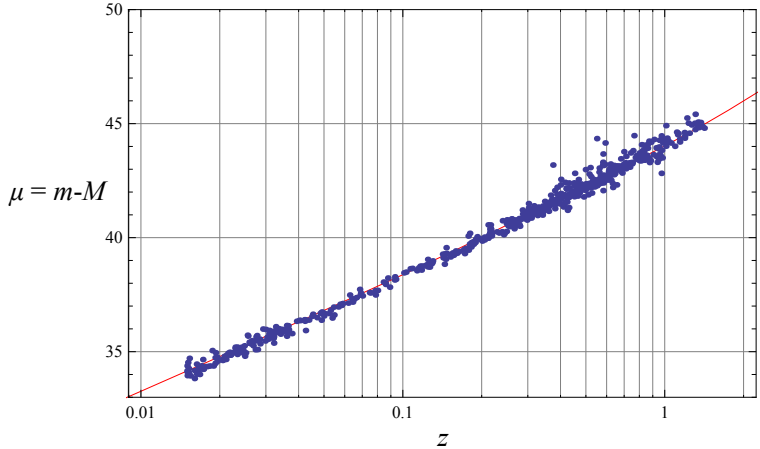}

\caption{\label{fig:mue-z-A}Dots: Observations of distance moduli $\mu=m-M$
and redshifts $z$ for Type Ia supernovae. $M$ is the absolute magnitude;
it is the same for all Type Ia supernovae and it has the value -19.\newline
Graph: The $(\mu,z)$-plot for our model of the universe fits well
to observed values.}
\end{figure}

The currently popular cosmology%
\footnote{See S. Weinberg \cite{Weinberg2008}, or R.Chavez \cite{Chavez2014}
for instructive presentations.%
} is much influenced by the need to understand or describe the observed
relation between the apparent magnitude $m$ of Type Ia supernovae
and the redshift $z$ of the light received from them. The observations,
reported by Riess et al., \cite{Riess2007}, Amanullah et al. \cite{Amanullah2010},
Hicken et al. \cite{Hicken2009} and others, compiled in the Union
2.0 and 2.1 catalogues \cite{Suzuki2012}, are reproduced in the diagram
of Fig. \ref{fig:mue-z-A} as dots. The diagram is often called the
Hubble diagram, because in some rough manner $m$ determines the distance
of the light source and $z$ represents its speed; and Hubble \cite{Hubble1929}
was first to conjecture a relation between speed and distance. That
conjecture gave rise to the notion of an expanding universe%
\footnote{R.P. Kirshner \cite{Kirshner2004} gives an instructive review about
Hubble\'{ }s law and the Hubble diagram. %
}. The need to understand the ($m$,$z$)-correlation has led cosmologists
to a revival of the cosmological constant%
\footnote{$\Lambda$ was first introduced by Einstein \cite{Einstein1922},
see also \cite{Einstein1922b}. Later the concept of a cosmological
constant was largely given up, not least by Einstein himself. %
} $\Lambda$, -- interpreted as a hypothetical repulsive \textquotedbl{}dark
energy\textquotedbl{} --, and the prediction of an accelerated expansion
of the universe. These notions are embedded in the concept of a four-dimensional
Euclidean space with a three-dimensional surface, either spherical
or flat or hyperspherical, and endowed with a Robertson-Walker metric.
The three-dimensional surface is supposed to be homogeneously filled
with matter, a strongly simplifying assumption that reduces the Einstein
equations of gravitation to a simple set of ordinary differential
equations. Friedman \cite{Friedman1922,Friedmann1924} has solved
the equations and accordingly this theory is often called the FRW-cosmology.
The study of the cosmic radiation background has led FRW-cosmologists
to the assumption that the universe is in fact flat. And if this is
the case, the Hubble diagram requires that the dark energy makes up
ca. $70\%$ of the mass of the universe (!) (see \cite{Weinberg2008},
p.48).

\subsection*{Present model}

In the present paper we investigate a less arcane model: We view the
universe as a sphere of the time- dependent radius $R(t)$ filled
with matter and floating freely in infinite empty space%
\footnote{Einstein in \cite{Einstein1922b} spoke of \textquotedbl{}an island
which floats in infinite empty space\textquotedbl{} and he rejects
the notion. We do not; rather we exploit it. %
}. The matter is supposed to be distributed isotropically with respect
to the center of the sphere and it is considered as a dust gas so
that pressure and internal energy are negligible and temperature plays
no role. Cosmic radiation is neglected as well and thus the only agent
dictating the motion of the dust gas is the gravitational effect on
it.

The motion is governed by the Einstein equations of gravitation, --
without the cosmological constant $\Lambda$. We solve these equations
for appropriate boundary conditions, -- in the center, at infinity,
and at $R(t)$ --, and for appropriate initial conditions at some
time $t_{i}$ in the past, when the expansion observed by Hubble is
already in progress. The initial conditions are choosen so as to provide
a good fit between the \textit{calculated} $\mu(z)$-curve and the
\textit{observed} data, see Fig. \ref{fig:mue-z-A}; the method is
known in mathematics as the solution of an \textit{inverse problem}.
From the initial time onwards we follow the continued expansion for
$t>t_{i}$ and, calculating backwards for $t<t_{i}$, we see that
the universe has emerged from a sphere with the Schwarzschild radius.

We come to the conclusion that the observed $\mu(z)$-data are compatible
with different solutions of the Einstein equations,-- always without
$\Lambda$. That is to say that different initial data lead to the
same $\mu(z)$-curve. We exhibit two solutions explicitly and none
of them shows an accelerated expansion nor do they require dark energy.
The non-uniqeness of the solution of our inverse problem means that
more astronomical observations are required to find a unique solution.
Which additional observations are feasible? And which solutions of
the Einstein equations describe them well? Those questions represent
the cosmological conundrum.

In our paper we choose Schwarzschild coordinates $\left(t,r,\vartheta,\varphi\right)$
for \textit{calculations}, i.e. for the formulation and reformulation
and solution of the Einstein equations. In these coordinates we have
a fairly simple diagonal metric tensor with only the radial and the
temporal component as initially unknown functions of $t$ and $r$,
while the angular components are like those in Euclidean space. For
the \textit{interpretation} of the results we replace the Schwarzschild
coordinate time $t$ by the proper time of the observer in the center
of the sphere.

It is true that in our model of the universe we, the observers of
luminosities and redshifts, are situated in the center of the spherical
dust gas or close to it. This violates the \textit{cosmological principle},
according to which we should not occupy a privileged place in the
cosmos. We do accept that violation because we believe that the privileged
place is less counter-intuitive than is a hypothetical dark energy.
Besides in Chapter 6 we conjecture that the position of our observer
need not necessarily be restricted to the central \textit{point} of
the sphere, rather it may lie within a fairly extensive \textit{sphere}.

\section{Einstein equations}

\subsection{General form}

The Einstein equations read (without cosmological constant)
\begin{equation}
G_{\alpha\beta}=8\pi\frac{\gamma}{c^{4}}T_{\alpha\beta},\label{eq:Einstein-equations}
\end{equation}
where $G_{\alpha\beta}=R_{\alpha\beta}-\frac{1}{2}Rg_{\alpha\beta}$
is the Einstein tensor and $T_{\alpha\beta}$ is the energy-momentum
tensor of the matter. $\gamma$ is the gravitational constant. $R_{\alpha\beta}$
is the Ricci tensor and $R=g^{\alpha\beta}R_{\alpha\beta}$ is the
Ricci scalar. The Ricci tensor is related to the Christoffel symbols
$\Gamma_{\delta\varepsilon}^{\gamma}$ which in turn are related to
the metric tensor $g_{\mu\nu}$. We have%
\footnote{Greek indices run from 0 to 3, Latin ones from 1 to 3, such that $x^{0}=ct$,
$x^{b}=\left(r,\vartheta,\varphi\right)$. $g_{\alpha\beta}$ is the
metric tensor in space-time. $g_{\alpha\beta}$ and its inverse $g^{\alpha\beta}$
may be used to lower and raise indices in the usual manner.%
}
\begin{equation}
R_{\alpha\beta}=\frac{\partial\Gamma_{\alpha\lambda}^{\lambda}}{\partial x^{\beta}}-\frac{\partial\Gamma_{\alpha\beta}^{\lambda}}{\partial x^{\lambda}}+\Gamma_{\alpha\lambda}^{\gamma}\Gamma_{\beta\gamma}^{\lambda}-\Gamma_{\alpha\beta}^{\lambda}\Gamma_{\gamma\lambda}^{\gamma}\:\textrm{and}\:\Gamma_{\alpha\beta}^{\lambda}=\frac{1}{2}g^{\lambda\gamma}\left(\frac{\partial g_{\gamma\alpha}}{\partial x^{\beta}}+\frac{\partial g_{\gamma\beta}}{\partial x^{\alpha}}-\frac{\partial g_{\alpha\beta}}{\partial x^{\gamma}}\right).\label{eq:Ricci-tensor}
\end{equation}

\subsection{Metric tensor. Speed and trajectory of light.}

For the present purpose, -- the consideration of a radially expanding
sphere, isotropically filled with matter -- we assume that the metric
tensor has the form
\begin{equation}
g_{ij}=\left(\begin{array}{cccc}
Z\\
 & S\\
 &  & r^{2}\\
 &  &  & r^{2}\sin^{2}\vartheta
\end{array}\right).\label{eq:Metric}
\end{equation}
Space-time coordinates $\left(t,r,\vartheta,\varphi\right)$ for which
the metric tensor has this form are called Schwarzschild coordinates.
Oppenheimer \& Volkoff \cite{Oppenheimer1939} and Oppenheimer \&
Snyder \cite{Oppenheimer1939b} have used Schwarzschild coordinates
for the description of a neutron star. Here we use them for the description
of the universe as a whole: A spherical dust gas in infinite empty
space. A sphere of radius $r$ in Schwarzschild coordinates obviously
has the surface area $4\pi r^{2}$ just like in Euclidean space, see
also Misner et al \cite{Misner1973}. The components $Z$ and $S$
may be functions of $t$ and $r$, and we rely on the Einstein equations
to find the dependence. $Z(t,r)$ must be negative, because without
gravitation $Z$ must be equal to -1. $S$ must be equal to 1 in that
case.

Since $Z(t,r)$ and $S(t,r)$ are components of the metric tensor,
we feel justified to assume that they must be continuous and smooth
functions, apart possibly from singular points.

It follows that the infinitesimal distance element $ds^{2}=g_{\alpha\beta}dx^{\alpha}dx^{\beta}$
in space-time is given in terms of the coordinate increments $dt$
and $dr$ by%
\footnote{We drop the angular part of $ds^{2}$ for brevity; it is unimportant
for radial motion.%
}
\begin{equation}
ds^{2}=Zc^{2}dt^{2}+Sdr^{2}.\label{eq:ds_eqn}
\end{equation}
In a local and momentary Lorentz frame at rest with a particle of
matter we denote the coordinate increments of time and radial distance
by $d\theta$ and $d\varrho$ so that we have
\begin{equation}
ds^{2}=-c^{2}d\theta{}^{2}+d\varrho^{2}.\label{eq:ds_eqn_eigen}
\end{equation}
$d\theta$ and $d\varrho$ are called the increments of \textit{proper}
time and \textit{proper} distance associated with the particle.

Hence follows a relation between $dt$ and $d\theta$ for fixed $\varrho$
\begin{equation}
d\theta=\sqrt{|Z|}\sqrt{1-\frac{V^{2}}{c^{2}}\frac{S}{|Z|}}dt,\:\textrm{where}\: V=\left(\frac{\partial r}{\partial t}\right)_{\varrho}\label{eq:d_tau}
\end{equation}
is the velocity of a material particle in the $(t,r)$-system. Similarly
a relation between $dr$ and the increment $d\varrho$ of proper distance
at fixed $\theta$ reads%
\footnote{While (\ref{eq:d_tau}) is an obvious consequence of (\ref{eq:ds_eqn}),
(\ref{eq:ds_eqn_eigen}), the relation (\ref{eq:d_rho}) requires
a few intermediate lines of calculation with partial derivatives in
order to determine that $\left(\frac{\partial t}{\partial r}\right)_{\theta}=\frac{V}{c^{2}}\frac{S}{|Z|}$
holds.%
}
\begin{equation}
d\varrho=\sqrt{S}\sqrt{1-\frac{V^{2}}{c^{2}}\frac{S}{|Z|}}dr.\label{eq:d_rho}
\end{equation}

The trajectory of a radial light ray in the $(t,r)$-system is governed
by the equation $ds^{2}=0$ so that, by (\ref{eq:ds_eqn}), we have
for the velocity $c_{x}$ of light moving radially
\begin{equation}
\left(\frac{\partial r}{\partial t}\right)_{s}=c_{x}=\pm c\sqrt{\frac{|Z|}{S}}.\label{eq:drdt_s}
\end{equation}
$|c_{x}|$ denotes the speed of light in the $(t,r)$-system. Obviously,
from (\ref{eq:d_tau}), (\ref{eq:d_rho}) the speed $V$ of a material
particle is bound by the speed of light in the $(t,r)$-system
\begin{equation}
|V|\leq c_{x}=c\sqrt{\frac{|Z|}{S}}.\label{eq:V_lower-then_c}
\end{equation}

Later we shall be interested in the \textit{trajectory of a light
ray} that reaches the observer of our spherical universe in the center
of the sphere at the present time $t=0$, i.e. at the event $(t,r)=(0,0)$.
That trajectory -- denoted by $r_{T}(t)$ -- must be obtained by integration
from (\ref{eq:drdt_s}). Naturally we need explicit functions $Z(t,r)$
and $S(t,r)$ for the purpose. Therefore such a calculation has to
be postponed until we have obtained those functions from the Einstein
equations.

\subsection{Specific form of the Einstein tensor}

It is a cumbersome task to calculate the Christoffel symbols, hence
the Ricci tensor, and hence the Einstein tensor from the specific
form (\ref{eq:Metric}) of $g_{\alpha\beta}$. However, when the calculations
are done, it turns out that the Einstein tensor contains only four
essential non-zero components and they read
\begin{equation}
\begin{array}{ccl}
G_{00} & = & \frac{1}{r^{2}}\frac{Z}{S}\left(1-S-r\frac{S'}{S}\right)\\
G_{0r} & = & \frac{1}{r}\frac{1}{c^{2}}\frac{\dot{S}}{S}\\
G_{rr} & = & \frac{1}{r^{2}}\left(1-S+r\frac{Z'}{Z}\right)\\
G_{\vartheta\vartheta} & = & r^{2}\frac{1}{4c^{2}}\frac{1}{ZS}\left(2\ddot{S}-\frac{1}{S}\dot{S}^{2}-\frac{1}{Z}\dot{Z}\dot{S}+2c^{2}Z''-c^{2}\frac{1}{Z}Z'^{2}-\frac{1}{S}S'Z'-\frac{2c^{2}}{r}\left(Z'-\frac{Z}{S}S'\right)\right).
\end{array}\label{eq:Einstein-tensor}
\end{equation}
$\left(\right)'$ and $\left(\right)^{\bullet}$ denote partial derivatives
with respect to $r$ and $t$ respectively. $G_{\varphi\varphi}$
is not zero, but it is equal to $G_{\vartheta\vartheta}$ to within
a factor $r^{2}\sin^{2}\vartheta$ and may therefore be ignored, because
the components $T_{\vartheta\vartheta}$ and $T_{\varphi\varphi}$
of the energy-momentum tensor both vanish in our case as we shall
presently see.

\subsection{Energy-momentum tensor for a dust gas}

We assume that the matter in the sphere is a dust gas. This means
that there are no viscosities and, in addition, that the pressure
may be neglected. If that is true, the energy-momentum tensor is given
by
\begin{equation}
T_{\alpha\beta}=\sigma U_{\alpha}U_{\beta},\label{eq:energy-momentum-tensor}
\end{equation}
where $\sigma$ is a scalar rest-mass density of the matter. The idea
is, of course, that the matter in the universe is \textquotedbl{}smeared
out\textquotedbl{} to form a continuum; each point is then occupied
by a \textquotedbl{}particle\textquotedbl{} of matter in the sense
of continuum mechanics.

$U_{\alpha}$ are the covariant components of the four-velocity and
$U^{\alpha}=g^{\alpha\beta}U_{\beta}$ are its contravariant components.
For radial expansion we have
\begin{equation}
U^{\alpha}=\frac{dx^{\alpha}}{d\theta}=\frac{dx^{\alpha}}{dt}\frac{dt}{d\theta}=\left(\begin{array}{c}
c\\
V\\
0\\
0
\end{array}\right)\frac{dt}{d\theta}.\label{eq:four-velocity}
\end{equation}
$V=\left(\frac{\partial r}{\partial t}\right)_{\varrho}$ is the radial
velocity of the gas as before, and $\theta$ is the proper-time. We
use (\ref{eq:d_tau}) and (\ref{eq:drdt_s}) to write (\ref{eq:four-velocity})
in the form 
\begin{equation}
U^{\alpha}=\left(\begin{array}{c}
c\\
V\\
0\\
0
\end{array}\right)\frac{\frac{1}{\sqrt{|Z|}}}{\sqrt{1-\frac{V^{2}}{c_{x}^{2}}}}\:\textrm{and therefore}\: U_{\alpha}=\left(\begin{array}{c}
-c|Z|\\
VS\\
0\\
0
\end{array}\right)\frac{\frac{1}{\sqrt{|Z|}}}{\sqrt{1-\frac{V^{2}}{c_{x}^{2}}}}.\label{eq:four-velocity-2}
\end{equation}
Note that the $r$-component $\left(\frac{\partial r}{\partial\theta}\right)_{\varrho}$
of the 4-velocity equals
\[
\left(\frac{\partial r}{\partial\theta}\right)_{\varrho}=\frac{1}{\sqrt{|Z|}}\frac{V}{\sqrt{1-\frac{V^{2}}{c_{x}^{2}}}}
\]
so that $|\left(\frac{\partial r}{\partial\theta}\right)_{\varrho}|$
varies between 0 and $\infty$ when $|V|$ varies between 0 and $|c_{x}|$,
the speed of light.

By (\ref{eq:energy-momentum-tensor}), (\ref{eq:four-velocity-2})
the energy-momentum tensor assumes the form
\begin{equation}
T_{\alpha\beta}=\left(\begin{array}{cccc}
\frac{\sigma c^{2}|Z|}{1-\frac{V^{2}}{c_{x}^{2}}} & \frac{-\sigma cVS}{1-\frac{V^{2}}{c_{x}^{2}}} & 0 & 0\\
\\
\frac{-\sigma cVS}{1-\frac{V^{2}}{c_{x}^{2}}} & \frac{\sigma V^{2}\frac{S^{2}}{|Z|}}{1-\frac{V^{2}}{c_{x}^{2}}} & 0 & 0\\
0 & 0 & 0 & 0\\
0 & 0 & 0 & 0
\end{array}\right).\label{eq:energy-momentum-tensor-2}
\end{equation}

\subsection{Specific form of the Einstein equations}

Elimination of $G_{\alpha\beta}$ and $T_{\alpha\beta}$ between (\ref{eq:Einstein-equations}),
(\ref{eq:Einstein-tensor}), and (\ref{eq:energy-momentum-tensor-2})
gives the specific form of the Einstein equations for the radial expansion
of an isotropic sphere filled by a dust gas. We obtain
\begin{equation}
\begin{array}{ccl}
S' & = & \frac{1}{r}S(1-S)+\frac{2}{r}\frac{\gamma}{c^{2}}m'S^{2},\:\textrm{where}\: m'\equiv\frac{\sigma4\pi r^{2}}{1-\frac{V^{2}}{c_{x}^{2}}}\\
\dot{S} & = & -\frac{2}{r}\frac{\gamma}{c^{2}}m'VS^{2}\\
Z' & = & -\frac{1}{r}Z(1-S)-\frac{2}{r}\frac{\gamma}{c^{2}}m'S^{2}\frac{V^{2}}{c^{2}}\\
0 & = & -2\ddot{S}+\frac{1}{S}\dot{S}^{2}+\frac{1}{Z}\dot{Z}\dot{S}-2c^{2}Z''-c^{2}\frac{1}{Z}Z'^{2}+c^{2}\frac{1}{S}S'Z'-\frac{2c^{2}}{r}\left(Z'-\frac{Z}{S}S'\right).
\end{array}\label{eq:specific-form-Einstein-eqn}
\end{equation}
We rely on these four equations for the determination of the four
fields $Z(t,r)$, $S(t,r)$, $\sigma(t,r)$, and $V(t,r)$. First,
however, we rewrite these equations aiming for a set of equations
that can be recognized as a relativistic generalization of the Newtonian
-- non relativistic -- set appropriate for a self-gravitating dust
gas.

First of all we note that (\ref{eq:specific-form-Einstein-eqn})\textsubscript{1}
may be written in the form
\[
\left[r\left(\frac{1}{S}-1\right)\right]^{'}=-2\frac{\gamma}{c^{2}}m'
\]
or, by integration from 0 to $r$ with $m(t,r)=\int_{0}^{r}m'(t,\alpha)d\alpha$
\begin{equation}
S=\frac{1}{1-2\frac{\gamma}{c^{2}}\frac{m}{r}}.\label{eq:S_from_m_r}
\end{equation}
$m$ is the mass within a sphere of radius $r$. Indeed, we observe
that the integrability condition for $S$ implied by (\ref{eq:specific-form-Einstein-eqn})\textsubscript{1,2}
reads
\begin{equation}
\dot{m}'+\left(m'V\right)'=0.\label{eq:mass-balance}
\end{equation}
It follows that $m'$ is the density of a conserved quantity. Therefore
$m'$ is the \textit{mass density} of a spherical shell referred to
the thickness $dr$ of the shell. {[}$m'$ must not be confused with
the scalar rest-mass density which is denoted by $\sigma$, cf. (\ref{eq:specific-form-Einstein-eqn})\textsubscript{1}.{]}
Thus, by (\ref{eq:S_from_m_r}) the dependent field $S(t,r)$ may
be replaced by the more intuitively appealing field $m(t,r)$. This
field satisfies a very simple differential equation which we derive
as follows. We observe that (\ref{eq:specific-form-Einstein-eqn})\textsubscript{1}
multiplied by $V$ and added to (\ref{eq:specific-form-Einstein-eqn})\textsubscript{2}
provides the relation
\[
\left[\ln\frac{S}{S-1}\right]^{\centerdot}+V\left[\ln\frac{S}{S-1}\right]^{'}=-\frac{V}{r},\:\textrm{or by (\ref{eq:S_from_m_r})}
\]
\[
\left[\ln\frac{m}{r}\right]^{\centerdot}+V\left[\ln\frac{m}{r}\right]^{'}=-\frac{V}{r},\:\textrm{or, finally}
\]
\[
\dot{m}+Vm'=0.
\]
It follows that the mass inside a spherical surface of radius $r$
is constant for the observer moving with that surface. That result
is eminently plausible, of course, and we write it as
\begin{equation}
\frac{dm}{dt}=0,\:\textrm{where}\:\frac{d\left(\right)}{dt}=\left(\right)^{\centerdot}+V\left(\right)'.\label{eq:mass-balance-2}
\end{equation}

The formidable equation (\ref{eq:specific-form-Einstein-eqn})\textsubscript{4}
may be given a somewhat more intuitively appealing form by elimination
of $\ddot{S}$, $Z''$, $\dot{S}$, $S'$ and by use of (\ref{eq:S_from_m_r}),
(\ref{eq:mass-balance}). We obtain after some calculation
\begin{equation}
\frac{dV}{dt}-\frac{1}{2}\frac{V}{Z}\frac{dZ}{dt}=\gamma\frac{m}{r^{2}}\left(Z+2\frac{V^{2}}{c^{2}}\frac{1}{1-\frac{2\gamma}{c^{2}}\frac{m}{r}}\right).\label{eq:momentum-balance}
\end{equation}
\newline  

\footnotesize This equation may be seen as the relativistic generalization
of the equation of motion of a self-gravitating dust gas. Indeed,
in the non-relativistic case, i.e. when $Z=-1$ and $V^{2}/c^{2}\approx0$,
we obtain from (\ref{eq:momentum-balance})
\[
\frac{dV}{dt}=-\gamma\frac{m}{r^{2}},
\]
which is the $r$-component of the non-relativistic equation of motion
with the classical attractive gravitational acceleration on the right-hand-side.
If an expansion is in progress, so that $V>0$ for all $r>0$ holds,
that term decelerates the expansion. Note however, that in (\ref{eq:momentum-balance})
the second term on the right-hand-side is positive, i.e. that term
causes an accelerated expansion. In Section 5.5 the effect of that
term is made explicit after the Einstein equations have been solved
and the fields $Z(t,r)$, $m(t,r)$, $\sigma(t,r)$, and $V(t,r)$
have been determined. Note also that the second term on the right-hand-side
of (\ref{eq:momentum-balance}) becomes singular when the sphere has
the radius $r_{S}=2\gamma M/c^{2}$ which is the Schwarzschild radius
for the total mass $M$.\normalsize\newline  

Finally we may eliminate the acceleration $dV/dt$ from (\ref{eq:momentum-balance})
by use of (\ref{eq:specific-form-Einstein-eqn})\textsubscript{1},
the definition of $m'$. Making use of (\ref{eq:mass-balance}), (\ref{eq:mass-balance-2})
we thus obtain
\begin{equation}
\frac{d\sigma}{dt}+\sigma V'+2\sigma\frac{V}{r}=\frac{2\gamma}{c^{2}}\frac{m}{r^{2}}\frac{1}{1-\frac{2\gamma}{c^{2}}\frac{m}{r}}\sigma V\label{eq:rho-DGL}
\end{equation}
or
\begin{equation}
\left(\sigma r^{2}\right)^{\bullet}+\left(\sigma r^{2}V\right)^{'}=\frac{2\gamma}{c^{2}}m\frac{1}{1-\frac{2\gamma}{c^{2}}\frac{m}{r}}\sigma V.\label{eq:rho-DGL_2}
\end{equation}
\newline  

\footnotesize This is clearly the relativistic generalization of
the non-relativistic law of mass-conservation, since the right-hand-side
is negligible in that case. Note that non-relativistically there is
no essential difference between $\sigma r^{2}$ and the mass density
$m'$.\normalsize\newline  

In summary we may now say that so far the original four Einstein equations
are replaced by the four equations
\begin{equation}
\begin{array}{lcl}
\textrm{\textrm{From (\ref{eq:specific-form-Einstein-eqn}\ensuremath{)_{1}}with (\ref{eq:S_from_m_r}):}} &  & \frac{V^{2}}{c^{2}}=Z\left(1-\frac{2\gamma}{c^{2}}\frac{m}{r}\right)\frac{\sigma4\pi r^{2}-m'}{m'}\\
\textrm{From (\ref{eq:specific-form-Einstein-eqn}\ensuremath{)_{3}}with (\ref{eq:S_from_m_r}) and (\ref{eq:specific-form-Einstein-eqn}\ensuremath{)_{1}}:} &  & \frac{Z'}{Z}=\frac{\frac{2\gamma}{c^{2}}\frac{m}{r^{2}}}{1-\frac{2\gamma}{c^{2}}\frac{m}{r^{2}}}\left(1-\frac{\sigma4\pi r^{2}-m'}{\frac{m}{r}}\right)\\
\textrm{From (\ref{eq:specific-form-Einstein-eqn}\ensuremath{)_{1}}and (\ref{eq:specific-form-Einstein-eqn}\ensuremath{)_{2}}:} &  & \frac{dm}{dt}=0\\
\textrm{From (\ref{eq:specific-form-Einstein-eqn}\ensuremath{)_{4}}:} &  & \frac{dV}{dt}-\frac{1}{2}\frac{V}{Z}\frac{dZ}{dt}=\gamma\frac{m}{r^{2}}\left(Z+2\frac{V^{2}}{c^{2}}\frac{1}{1-\frac{2\gamma}{c^{2}}\frac{m}{r}}\right).
\end{array}\label{eq:four_field_eqns}
\end{equation}

Alternatively -- replacing (\ref{eq:four_field_eqns})\textsubscript{1}
by its corollary (\ref{eq:rho-DGL}) -- we may write this set of equations
as
\begin{equation}
\begin{array}{l}
\frac{d\sigma}{dt}+\sigma V'+2\sigma\frac{V}{r}=\frac{2\gamma}{c^{2}}\frac{m}{r^{2}}\frac{1}{1-\frac{2\gamma}{c^{2}}\frac{m}{r}}\sigma V\\
\frac{Z'}{Z}=\frac{\frac{2\gamma}{c^{2}}\frac{m}{r^{2}}}{1-\frac{2\gamma}{c^{2}}\frac{m}{r^{2}}}\left(1-\frac{\sigma4\pi r^{2}-m'}{\frac{m}{r}}\right)\\
\frac{dm}{dt}=0\\
\frac{dV}{dt}-\frac{1}{2}\frac{V}{Z}\frac{dZ}{dt}=\gamma\frac{m}{r^{2}}\left(Z+2\frac{V^{2}}{c^{2}}\frac{1}{1-\frac{2\gamma}{c^{2}}\frac{m}{r}}\right).
\end{array}\label{eq:four_field_eqns-2}
\end{equation}
Yet another arrangement of these equations is as follows
\begin{equation}
\begin{array}{l}
\frac{V^{2}}{c^{2}}=Z\left(1-\frac{2\gamma}{c^{2}}\frac{m}{r}\right)\frac{\sigma4\pi r^{2}-m'}{m'}\\
\frac{Z'}{Z}=\frac{\frac{2\gamma}{c^{2}}\frac{m}{r^{2}}}{1-\frac{2\gamma}{c^{2}}\frac{m}{r^{2}}}\left(1-\frac{\sigma4\pi r^{2}-m'}{\frac{m}{r}}\right)\\
\frac{dm}{dt}=0\\
\frac{d\sigma}{dt}+\sigma V'+2\sigma\frac{V}{r}=\frac{2\gamma}{c^{2}}\frac{m}{r^{2}}\frac{1}{1-\frac{2\gamma}{c^{2}}\frac{m}{r}}\sigma V.
\end{array}\label{eq:four_field_eqns-3}
\end{equation}
This results from replacing (\ref{eq:four_field_eqns})\textsubscript{4}
by its corollary (\ref{eq:rho-DGL}). The three sets (\ref{eq:four_field_eqns})
through (\ref{eq:four_field_eqns-3}) are equivalent; their different
forms are needed here for arguments of interpretation and in the numerical
solution of the set.

\subsection{Non-dimensional variables}

For the subsequent numerical calculations we introduce dimensionless
dependent and independent variables based on the speed of light $c$,
the total mass $M$ of the universe and its present radius $R(0)$.
Note that both $M$ and $R(0)$ are \textit{a priori} unknown; they
must be chosen so as to be compatible with the $\mu(z)$-curve. We
define
\begin{equation}
\hat{r}=\frac{r}{R(0)},\;\hat{t}=\frac{ct}{R(0)},\;\hat{R}=\frac{R}{R(0)},\;\hat{\sigma}=\frac{\sigma\frac{4\pi}{3}R^{3}(0)}{M},\;\hat{m}=\frac{m}{M},\;\hat{V}=\frac{V}{c},\;\hat{c}_{x}=\frac{c_{x}}{c}.\label{eq:nondim-var}
\end{equation}
Also we introduce the dimensionless parameter
\begin{equation}
Q=\frac{2\gamma}{c^{2}}\frac{M}{R(0)},\label{eq:Q_eqn}
\end{equation}
which represents the Schwarzschild radius of the universe referred
to its present radius $R(0)$.

\section{External solution for \textit{S}(\textit{t},\textit{r}) and \textit{Z}(\textit{t},\textit{r})}

In the space outside the matter-filled sphere the rest mass density
$\sigma$ is equal to zero so that (\ref{eq:specific-form-Einstein-eqn})\textsubscript{1}
may be integrated to give $S=K(t)/(1-K(t)/r)$. $K(t)$ is a constant
of integration whose value is given by (\ref{eq:S_from_m_r}) as $\left(2\gamma/c^{2}\right)M$
, because at the surface of the sphere $S$ must be continuous and
equal to the value given by (\ref{eq:S_from_m_r}) with $m=M\equiv m(t,R(t))$
which is the constant total mass. Thus we have for the external solution
of $S$
\begin{equation}
S_{ext}=\frac{1}{1-\frac{2\gamma}{c^{2}}\frac{M}{r}}\:\textrm{for}\: r>R(t),\:\textrm{or}\: S_{ext}=\frac{1}{1-Q\frac{1}{\hat{r}}}\:\textrm{for}\:\hat{r}>\hat{R}(\hat{t}).\label{eq:S_ext}
\end{equation}
If we require $S(t,r)$ to be continuous \textit{and} smooth at the
surface, we must obviously have $m'(t,R(t))=0$.

$Z_{ext}$, the outer solution of $Z(t,r)$, follows by insertion
of (\ref{eq:S_ext}) into (\ref{eq:four_field_eqns})\textsubscript{2},
and integration: $Z_{ext}=k(t)\left(1-\frac{2\gamma}{c^{2}}\frac{M}{r}\right)$,
where $k(t)$ is a constant of integration whose value must be equal
to $-1$, so that $Z\rightarrow-1$ for $r\rightarrow\infty$, where
gravitation is negligible. Thus we have for the outer solution of
$Z$
\begin{equation}
Z_{ext}=-1+\frac{2\gamma}{c^{2}}\frac{M}{r}\:\textrm{for}\: r>R(t)\;\textrm{or}\; Z_{ext}=-1+Q\frac{1}{\hat{r}}\:\textrm{for}\:\hat{r}>\hat{R}(\hat{t}).\label{eq:Z_ext}
\end{equation}

Neither $S_{ext}$ nor $Z_{ext}$ depend on time. And obviously $S_{ext}$
is singular at
\begin{equation}
r=r_{S}\equiv\frac{2\gamma}{c^{2}}M,\;\textrm{or}\;\hat{r}=\hat{r}_{S}=Q,
\end{equation}
and $Z$ is equal to zero at that radius which is known as the Schwarzschild
radius. Therefore the metric (\ref{eq:Metric}) loses its meaning
there.

\section{Internal solutions}

\subsection{Distance modulus and redshift}

The distance modulus $\mu$ and the redshift $z$ of Type Ia supernovae
represent reliable information from which we may infer some knowledge
about the structure and motion of the universe, see Fig \ref{fig:mue-z-A}.
That information comes to us at the event $(t,r)=(0,0)$ through the
light of stars which was emitted in the past at events $(t,r_{T}(t))$
with $t<0$ and with $0<r_{T}(t)<R(t)$. $r_{T}(t)$ is the trajectory
of light introduced in Section 2.2. $\mu$ and $z$ depend on the
values of the fields $\sigma$, $m$, $Z$ and $V$ at the event $(t,r_{T}(t))$
of emission and for our model that dependence is given by the functions
\begin{equation}
\begin{array}{c}
z(t)=z(t,r_{T}(t))=\sqrt{\frac{S(0,0)}{S(t,r_{T}(t))}}\sqrt{\frac{1+\frac{V(t,r_{T}(t))}{\underset{x}{c}(t,r_{T}(t))}}{1-\frac{V(t,r_{T}(t))}{\underset{x}{c}(t,r_{T}(t))}}}-1,\\
\mu(t)=\mu(t,r_{T}(t))=5\log_{10}\left[\frac{1}{10\textrm{pc}}r_{T}(t)\sqrt{\frac{S(t,r_{T}(t))}{S(0,0)}}(1+z)^{2}\right].
\end{array}\label{eq:z_und_mue_von_t}
\end{equation}
These equations are derived in the Appendix, see (\ref{eq:z_eqn})
and (\ref{eq:mue_eqn})\textsubscript{2} respectively. Therefore
our model of the universe -- if it is valid -- must have solutions
$\sigma(t,r)$, $m(t,r)$ (or $S(t,r)$), $Z(t,r)$, and $V(t,r)$
which are consistent with the observed luminosities and redshifts.
Such solutions will be found in this chapter.

\subsection{Initial conditions $\hat{\sigma}(\hat{t}_{i},\hat{r})$ and $\hat{m}(\hat{t}_{i},\hat{r})$}

We take the initial time as some $t_{i}<0$ and our plan is to calculate
forward from there into the range $t>t_{i}$ and backwards into the
range $t<t_{i}$. Inspection of (\ref{eq:four_field_eqns})\textsubscript{1,2}
shows that we need to impose initial conditions $\hat{\sigma}(\hat{t}_{i},\hat{r})$
and $\hat{m}(\hat{t}_{i},\hat{r})$ in order to determine $\hat{Z}(\hat{t}_{i},\hat{r})$
and $\hat{V}(\hat{t}_{i},\hat{r})$, the former by integration of
(\ref{eq:four_field_eqns})\textsubscript{2}. The constant of integration
may be found by the assumed continuity of $Z$ at the surface $R(t_{i})$
of the initial sphere. But smoothness of $Z$ -- also assumed -- needs
restrictions on the initial functions $\hat{\sigma}(\hat{t}_{i},\hat{r})$
and $\hat{m}(\hat{t}_{i},\hat{r})$; by (\ref{eq:four_field_eqns})\textsubscript{1,2}
it requires that $\hat{\sigma}(\hat{t}_{i},\hat{R}(t_{i}))$ and $\left.\frac{\partial\hat{m}}{\partial\hat{r}}\right|_{\left(\hat{t}_{i},\hat{R}(t_{i})\right)}$
both vanish. Also, obviously, we must have $\hat{m}(\hat{t}_{i},0)=0$
and $\hat{m}(\hat{t}_{i},\hat{R}(t_{i}))=1$ and, for reasons of symmetry,
$\left.\frac{\partial\hat{\sigma}}{\partial\hat{r}}\right|_{(\hat{t},0)}=0$
and $\hat{V}(\hat{t},0)=0$. Finally, by (\ref{eq:four_field_eqns})\textsubscript{1},
$\left.\frac{\partial\hat{m}}{\partial\hat{r}}\right|_{(\hat{t},\hat{r})}$
must be bigger than $3\hat{\sigma}(\hat{t},\hat{r})4\pi\hat{r}^{2}$
for all times and all radii, lest $V^{2}$ be negative.

Apart from these conditions $\hat{\sigma}(\hat{t}_{i},\hat{r})$ and
$\hat{m}(\hat{t}_{i},\hat{r})$ are arbitrary or, at least, there
is nothing in the theory that would restrict their generality. Such
a circumstance is commonplace in the solution of partial differential
equations, the situation merely needs the input of observed -- or
otherwise given -- initial data. Here, however, in an application
of partial differential equations to the universe, there is a problem:
Observations are either impossible or insufficient. For instance an
observer at the event $(t_{i},0)$ cannot possibly measure $\sigma(t_{i},r)$,
because light with its finite speed needs time to travel from $r$
to the center at $r=0$.

In this situation we have to determine the initial conditions $\hat{\sigma}(\hat{t}_{i},\hat{r})$
and $\hat{m}(\hat{t}_{i},\hat{r})$ in a complex argumentative loop
as follows. We \textit{assume} trial functions $\hat{\sigma}_{trial}(\hat{t}_{i},\hat{r})$,
$\hat{m}_{trial}(\hat{t}_{i},\hat{r})$ and determine the corresponding
solution $\hat{\sigma}(\hat{t},\hat{r})$, $\hat{m}(\hat{t},\hat{r})$,
$\hat{Z}(\hat{t},\hat{r})$, $\hat{V}(\hat{t},\hat{r})$ of the Einstein
equations, hence the trajectory $r_{T}(t)$ and hence the distance
modulus $\mu(\hat{t},\hat{r}_{T}(\hat{t}))$ and the redshift $z(\hat{t},\hat{r}_{T}(\hat{t}))$
from (\ref{eq:z_und_mue_von_t}) and finally $\mu(z)$ by elimination
of $\hat{t}$. If the agreement with the observed $\mu(z)$-relation
is good, so were the trial functions. If not, we have to change the
trial functions and try again until we have a satisfactory agreement.

Such a procedure is known as the solution of an inverse problem, a
complex variant of the simple shooting method. It is a cumbersome
and time-consuming process. However, it can be automated to a certain
extent. Also the process may be abbreviated by an intelligent choice
of the trial functions and it may be directed by intermediate results.
If well-conducted, the process in the end amounts to the adjustment
of a few parameters to the observed data.

As specific trial functions for $\hat{\sigma}(\hat{t}_{i},\hat{r})$
and $\hat{m}(\hat{t}_{i},\hat{r})$ we choose
\begin{equation}
\begin{array}{l}
\hat{\sigma}_{trial}(\hat{t}_{i},\hat{r})=\hat{\sigma}(\hat{t}_{i},0)\left(1-\left(\frac{\hat{r}}{\hat{R}(\hat{t}_{i})}\right)^{2n}\right)\left(1-a\left(\frac{\hat{r}}{\hat{R}(\hat{t}_{i})}\right)^{2}\right),\;\textrm{with}\; n\geq2,\: a\leq1\\
\left.\frac{\partial\hat{m}_{trial}}{\partial\hat{r}}\right|_{(\hat{t}_{i},\hat{r})}=3\hat{\sigma}_{trial}(\hat{t}_{i},\hat{r})\hat{r}^{2}B(\hat{r}),\;\textrm{where}\; B(\hat{r})=\frac{1}{1-\left(b_{1}\hat{r}+b_{3}\hat{r}^{3}+b_{5}\hat{r}^{5}\right)^{2}}.
\end{array}\label{eq:AB_rho_m}
\end{equation}
$\hat{m}_{trial}(\hat{t}_{i},\hat{r})$ itself follows by integration
of (\ref{eq:AB_rho_m})\textsubscript{2} over $\hat{r}$ between
0 and $\hat{r}$ , since $\hat{m}(\hat{t}_{i},0)=0$ holds. And $\hat{\sigma}(\hat{t}_{i},0)$
may be calculated from the requirement that $\hat{m}(\hat{t}_{i},\hat{R}(\hat{t}_{i}))$
be equal to 1. Fairly obviously the trial functions obey all the conditions
we have required. In particular, the form of $B(\hat{r})$ ensures
that $\hat{V}^{2}\geq0$ holds. This leaves us with five parameters
in (\ref{eq:AB_rho_m}) to be determined in the laborious adaptive
process of solution of our inverse problem, namely $n$, $a$, $b_{1}$,
$b_{3}$, $b_{5}$. Also $Q$, the dimensionless Schwarzschild radius,
and $R(0)$, the present radius, enter into this problem. Two convenient
choices of the outer radius $\hat{R}(\hat{t}_{i})$ of the sphere
at two initial times $\hat{t}_{i}$ are $\hat{R}(\hat{t}_{i})=0.44$
and $\hat{R}(\hat{t}_{i})=0.25$. After the calculation the corresponding
times $t_{i}$ may be read off from Fig. \ref{fig:R_and_VR} as $t_{i}=-0.677$
and $t_{i}=-0.789$ respectively, because $\hat{R}(0)$ must be equal
to 1.

\begin{figure}[H]
\includegraphics[scale=0.8]{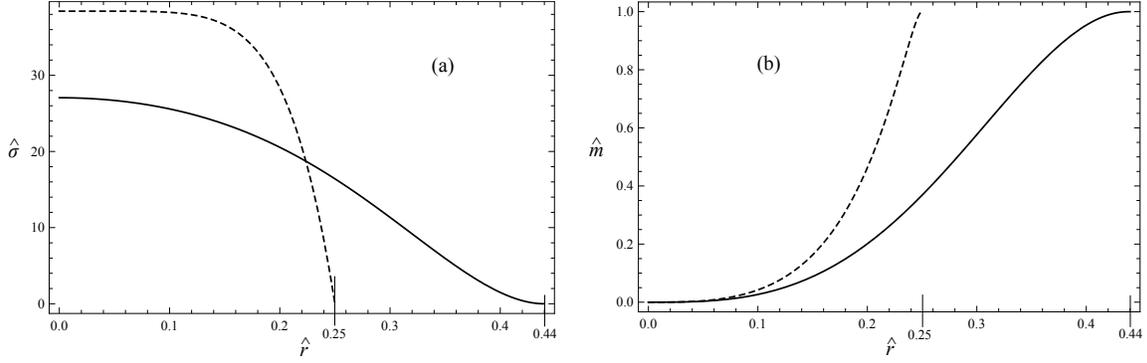}\caption{\label{fig:sigAB_mAB}Initial functions $\hat{\sigma}(\hat{t_{i}},\hat{r})$
and $\hat{m}(\hat{t}_{i},\hat{r})$ for the parameter choices A (solid)
and B (dashed).}
\end{figure}

We do not exhibit the adaptive process in detail. However, we anticipate
its result in order to be able to discuss the subsequent steps: As
it turns out, the triple $Q$, $n$, $a$ need not necessarily be
involved in the adaptation in order to obtain good results, only the
$b$\'{ }s, and $\hat{R}(0)$. We find that two choices of the parameters,
denoted by A and B, give nearly perfect results in the sense that
the calculated function $\mu(z)$ lies on the observed points in the
fashion demonstrated in Fig. \ref{fig:mue-z-A}. These \textquotedbl{}good\textquotedbl{}
choices are listed in Table \ref{tab:Two-good-choices}. And the corresponding
good initial functions $\hat{\sigma}(\hat{t}_{i},\hat{r})$ and $\hat{m}(\hat{t}_{i},\hat{r})$
are exhibited in Fig. \ref{fig:sigAB_mAB}.

\begin{table}[H]
\caption{\label{tab:Two-good-choices}Two \textquotedbl{}good\textquotedbl{}
parameter choices in the initial conditions for $\hat{\sigma}(\hat{t}_{i},\hat{r})$
and $\hat{m}(\hat{t}_{i},\hat{r})$ at time $\hat{t}_{i}$.}

\begin{tabular}{|c|c|c|c|c|c|c|c|}
\hline 
 & Q & n & a & $b_{1}$ & $b_{3}$ & $b_{5}$ & $R(0)$\tabularnewline
\hline 
\hline 
A & 0.11 & 2 & 1 & 1.77 & -0.94 & 17.65 & $10.79*10^{9}$ly\tabularnewline
\hline 
B & 0.015 & 3 & 0 & 3.9125 & 0 & 0 & $14.27*10^{9}$ly\tabularnewline
\hline 
\end{tabular}
\end{table}

The fact that there are \textit{two} \textquotedbl{}good\textquotedbl{}
sets of trial functions emphasizes the non-uniqueness of the solution
of our inverse problem. There are sure to be many more good solutions,
meaning that more astronomical observations -- in addition to $\mu$-
and $z$-measurements -- are needed to describe the universe properly.

\subsection{Initial functions for $Z(\hat{t}_{i},\hat{r})$ and $\hat{V}(\hat{t}_{i},\hat{r})$}

\begin{figure}[H]
\includegraphics[scale=0.8]{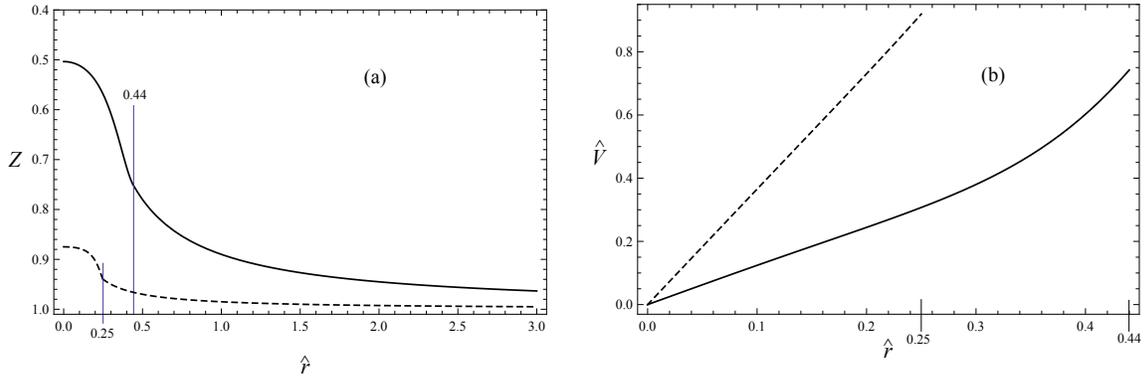}\caption{\label{fig:ZAB_VAB}Initial functions for $Z(\hat{t}_{i},\hat{r})$
and $V(\hat{t}_{i},\hat{r})$ for parameter choices A (solid) and
B (dashed). The graphs of $Z(\hat{t}_{i},\hat{r})$ combine the internal
solution with the external solution (\ref{eq:Z_ext})}
\end{figure}

Given $\hat{\sigma}(\hat{t}_{i},\hat{r})$ and $\hat{m}(\hat{t}_{i},\hat{r})$
we may now calculate the initial functions $Z(\hat{t}_{i},\hat{r})$
and $\hat{V}(\hat{t}_{i},\hat{r})$ from the Einstein equations (\ref{eq:four_field_eqns})\textsubscript{1,2}.
Integration of (\ref{eq:four_field_eqns})\textsubscript{2} provides
$Z(\hat{t}_{i},\hat{r})$ and subsequent insertion of this initial
function into (\ref{eq:four_field_eqns})\textsubscript{1} provides
$\hat{V}(\hat{t}_{i},\hat{r})$. The graphs are shown in Fig. \ref{fig:ZAB_VAB}.

\subsection{Step-wise solution for $\hat{\sigma}(\hat{t},\hat{r})$, $\hat{m}(\hat{t},\hat{r})$
, $\hat{V}(\hat{t},\hat{r}),$ $\hat{Z}(\hat{t},\hat{r})$ and $\hat{R}(\hat{t})$}

We proceed to find the solution of the Einstein equations (\ref{eq:four_field_eqns}),
or (\ref{eq:four_field_eqns-2}), or (\ref{eq:four_field_eqns-3}).
All functions depend on $(\hat{t},\hat{r})$.

The algebraic equation (\ref{eq:four_field_eqns-3})\textsubscript{1}
is only used to calculate the initial condition for $\hat{V}$. In
order to avoid the indeterminate expression $(3\hat{\sigma}\hat{r}^{2}-\hat{m}')/\hat{m}'$
in (\ref{eq:four_field_eqns-3})\textsubscript{1} at $\hat{r}=\hat{R}(\hat{t})$
we use the set (\ref{eq:four_field_eqns-2}) for the numerical solution
instead of (\ref{eq:four_field_eqns-3}). This set represents a coupled
system of three partial differential equations for $\hat{m}$, $\hat{\sigma}$,
$\hat{V}$ and one ordinary differential equation for $Z$.

For a very small time step $\Delta\hat{t}$ we replace $\hat{V}$
in the equation (\ref{eq:four_field_eqns-2})\textsubscript{3} by
the initial condition. The result is the simple Burgers equation
\[
\frac{\partial\hat{m}(\hat{t},\hat{r})}{\partial t}+\hat{V}(\hat{t}_{i},\hat{r})\frac{\partial\hat{m}(\hat{t},\hat{r})}{\partial\hat{r}}=0,
\]
in which $\hat{m}(\hat{t},\hat{r})$ is the only unknown. For the
initial condition of $\hat{m}$ we have $\hat{m}(\hat{t}_{i},\hat{r})$
by (\ref{eq:AB_rho_m})\textsubscript{2} on the interval $0\leq\hat{r}\leq\hat{R}(\hat{t}_{i})$.
The solution follows by the numerical method of characteristics on
the interval $0\leq\hat{r}\leq\hat{R}(\hat{t}_{i}+\Delta\hat{t})$.
The equation (\ref{eq:four_field_eqns-2})\textsubscript{1} will
also give a Burgers equation for $\hat{\sigma}$ by replacing $\hat{m}$
and $\hat{V}$ by their initial conditions. This equation is again
solved by the numerical method of characteristics.

Now we have $\hat{m}(\hat{t},\hat{r})$ and $\hat{\sigma}(\hat{t},\hat{r})$
and these functions are inserted in the equation (\ref{eq:four_field_eqns-2})\textsubscript{2}
for $Z$, which may be solved by a standard method for ordinary differential
equations.

At the end of the first time step we solve the partial differential
equation (\ref{eq:four_field_eqns-2})\textsubscript{4} for $\hat{V}$
in the same manner as the equation for $\hat{\sigma}$.

The solution $\hat{m}(\hat{t},\hat{r})$, $\hat{\sigma}(\hat{t},\hat{r})$,
$Z(\hat{t},\hat{r})$, $\hat{V}(\hat{t}_{i},\hat{r})$ on $\hat{t}_{i}\leq\hat{t}\leq\hat{t}_{i}+\Delta\hat{t}$
and $0\leq\hat{r}\leq\hat{R}(\hat{t}_{i}+\Delta\hat{t})$ is then
used as initial condition for the next time step.

\subsection{Summary of results}

\begin{figure}
\includegraphics[scale=0.65]{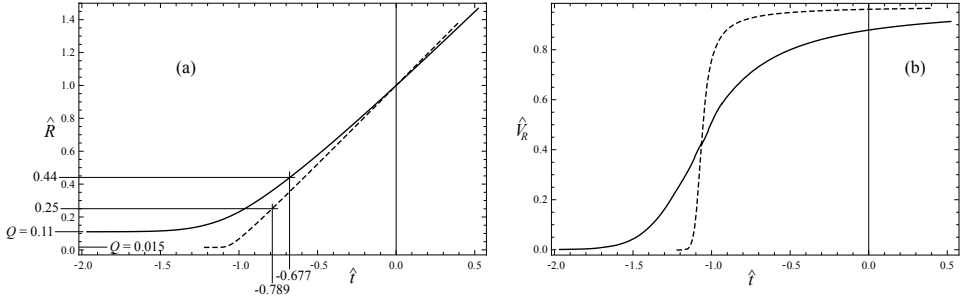}\caption{\label{fig:R_and_VR}$\hat{R}(\hat{t})$ and $\hat{V}_{R}(\hat{t})$
for parameter choices A (solid) and B (dashed)}
\end{figure}

First and foremost the solution of the Einstein equations provides
the radius of the universe as a function of time as shown in Fig.
\ref{fig:R_and_VR}a. The present time $t=0$ is identified as the
abscissa of the radius $\hat{R}=1$ . Thus we see that the time $t_{i}$,
which we have arbitrarily chosen as the initial time with $\hat{R}(\hat{t}_{i})=0.44$
for choice A is equal to $\hat{t}_{i}=-0.677$, while for $\hat{R}(\hat{t_{i}})=0.25$
in choice B it is equal $\hat{t}_{i}=-0.789$. $\hat{R}(\hat{t})$
grows indefinitely in the future toward an asymptotic slope. Also
inspection of the graph $\hat{R}(\hat{t})$ in the past reveals an
asymptotic approach of $\hat{R}(\hat{t})$ toward the Schwarzschild
radius $Q$ as $t$ tends to small values. Thus we conclude that the
expanding universe has emerged as a sphere with that radius. Fig.
\ref{fig:R_and_VR}b shows plots of $\hat{V}_{R}(\hat{t})$, the surface
velocity of the expanding sphere which results from differentiation
of $\hat{R}(\hat{t})$. It starts with zero velocity at small times
and continues to grow asymptotically toward $\hat{V}_{R}=1$. The
growth of $\hat{R}(\hat{t})$ and $\hat{V}_{R}(\hat{t})$ seems to
be accelerating, since the slope of $\hat{V}_{R}(\hat{t})$ is positive
and since the graph of $\hat{R}(\hat{t})$ is convex, i.e. it has
a positive curvature; at least $\hat{V}_{R}(\hat{t})$ does not stop
growing, as far as we can tell, even for large times. We come back
to this point in Chapter 5.

\begin{figure}
\includegraphics[scale=0.8]{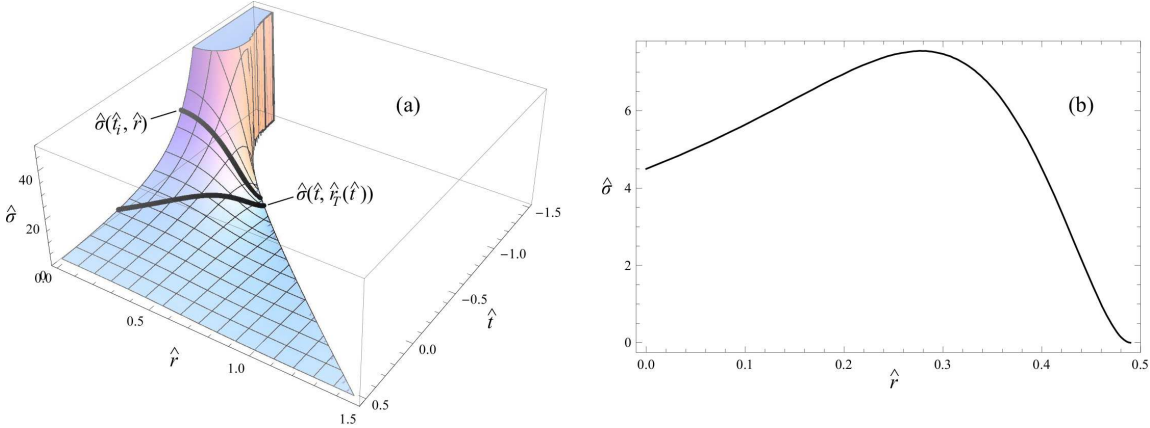}\caption{\label{fig:sig_3D_A}(a): $\hat{\sigma}(\hat{t},\hat{r})$. (b): $\hat{\sigma}(\hat{r})$
along the trajectory. Parameter choice A.}
\end{figure}

\begin{figure}
\includegraphics[scale=0.8]{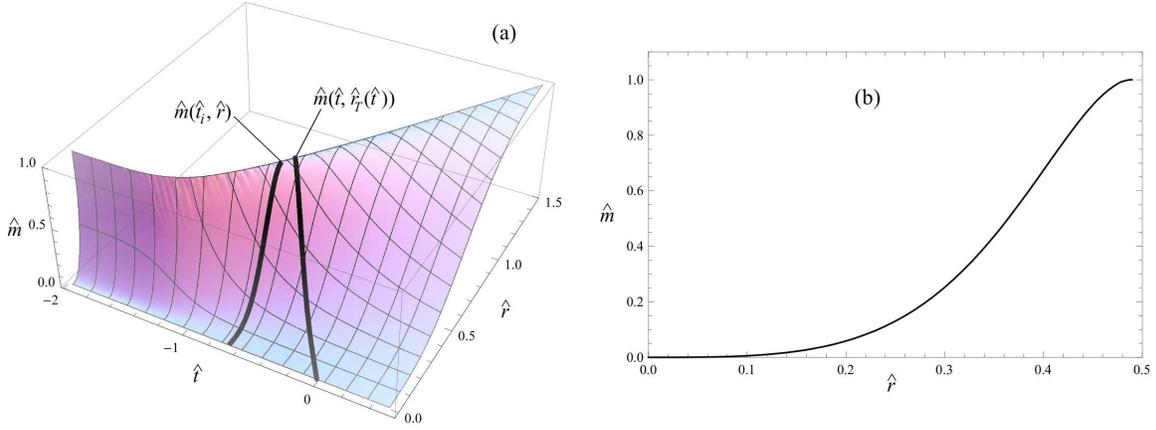}\caption{\label{fig:m_3D_A}(a): $\hat{m}(\hat{t},\hat{r})$. (b): $\hat{m}(\hat{r})$
along the trajectory. Parameter choice A.}
\end{figure}

\begin{figure}
\includegraphics[scale=0.8]{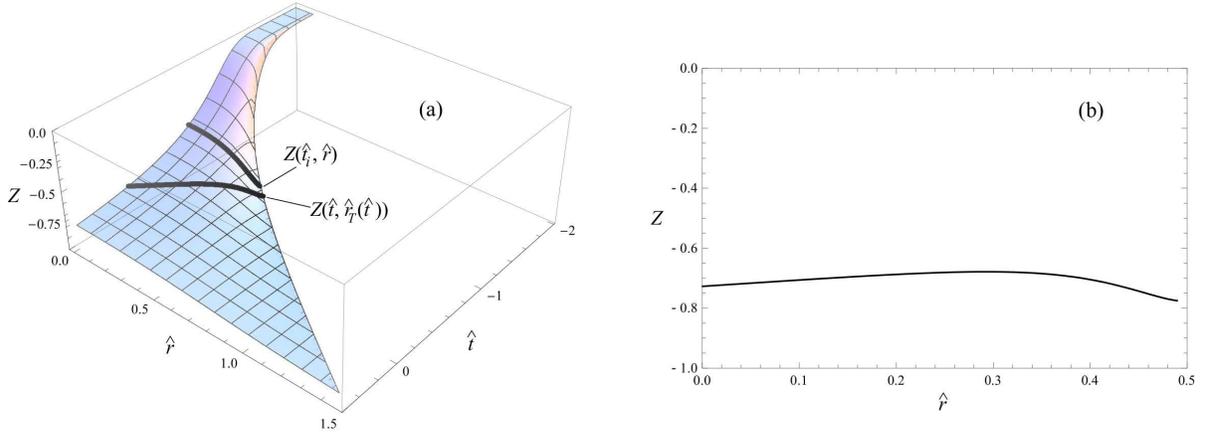}\caption{\label{fig:Z_3D_A}(a): $Z_{int}(\hat{t},\hat{r})$. (b): $Z(\hat{r})$
along the trajectory. Parameter choice A.}
\end{figure}

\begin{figure}
\includegraphics[scale=0.8]{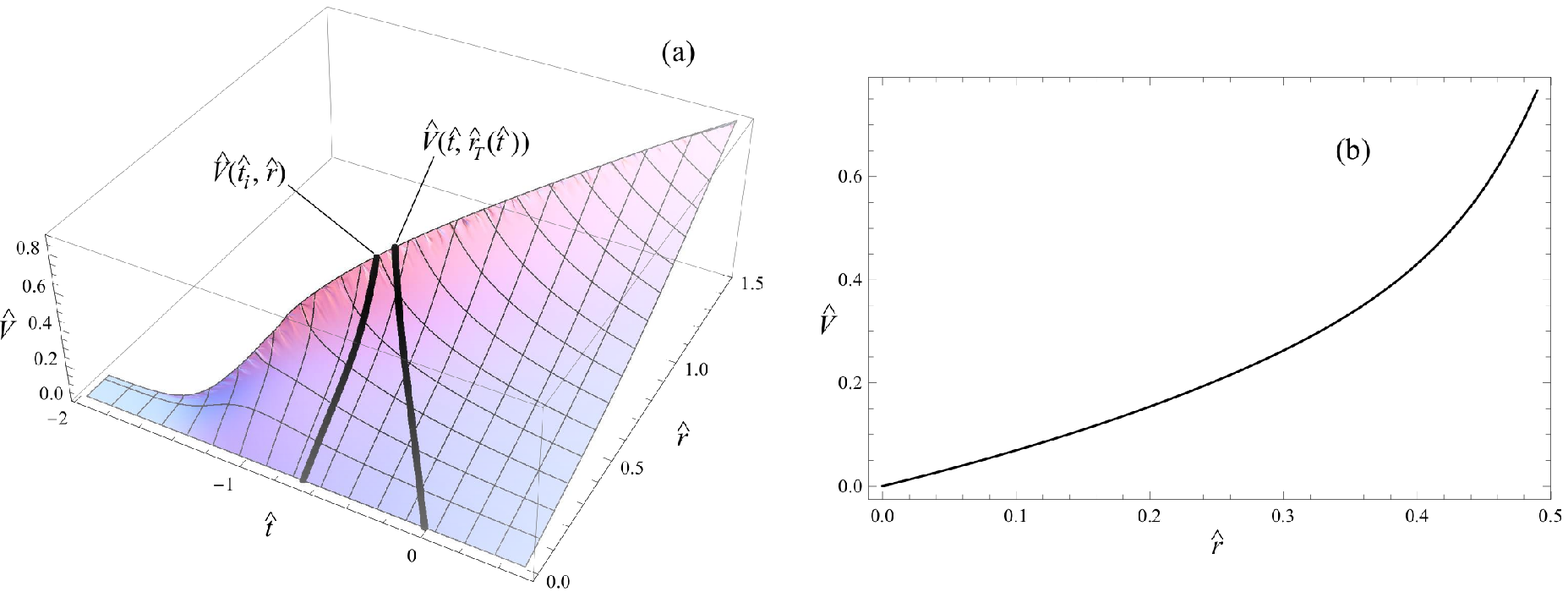}\caption{\label{fig:V_3D_A}(a): $\hat{V}(\hat{t},\hat{r})$. (b): $\hat{V}(\hat{r})$
along the trajectory. Parameter choice A.}
\end{figure}

The main results of our calculation are the fields $\hat{\sigma}(\hat{t},\hat{r})$
, $\hat{m}(\hat{t},\hat{r})$ , $\hat{V}(\hat{t},\hat{r})$ and $\hat{Z}(\hat{t},\hat{r})$.
Since they are fields on the $(t,r)$-plane, the best way to represent
them is by 3-D plots%
\footnote{An alternative is to show movies $\hat{\sigma}(\hat{t},\hat{r})$
, $\hat{m}(\hat{t},\hat{r})$ , $\hat{V}(\hat{t},\hat{r})$, $\hat{Z}(\hat{t},\hat{r})$
in time $t$ for $0\le\hat{r}\le\hat{R}(\hat{t})$. That presentation
will be used later in Chapter 6, at least for $\sigma(t,r)$ and $V(t,r)$,
of which we present screenshots for successive times.%
}. Such plots are shown in Fig. \ref{fig:sig_3D_A} through \ref{fig:V_3D_A}
for times $-1.5\le\hat{t}\le0.5$ or $-2.0\le\hat{t}\le0.5$ and in
the interval $0\le\hat{r}\le\hat{R}(\hat{t})$. The plots use different
perspectives, depending on whatever seems most appropriate to us.
All 3-D plots represent solutions for the parameter choice A.

\begin{figure}
\includegraphics[scale=0.8]{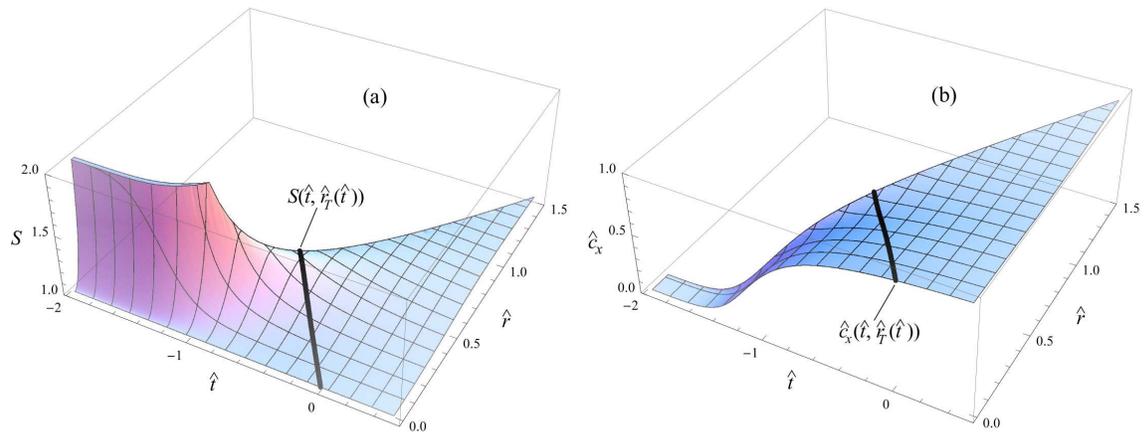}\caption{\label{fig:S_3D_cx_A}(a):$S(\hat{t},\hat{r})$. (b): $\hat{c}_{x}(\hat{t},\hat{r})$.
Parameter choice A}
\end{figure}

Additional important information furnished by the Einstein equations
are shown in Fig. \ref{fig:S_3D_cx_A}, which represent the fields
\begin{equation}
S(t,r)\;\textrm{and}\;\left(\frac{\partial r}{\partial t}\right)_{s}=c_{x}=c\sqrt{\frac{\left|Z(t,r)\right|}{S(t,r)}},\label{eq:cx_eqn}
\end{equation}
$S(t,r)$ follows from $m(t,r)$ by (\ref{eq:S_from_m_r}) and $c_{x}$,
the speed of light, is recalled from (\ref{eq:drdt_s}). The corresponding
3-D plots are shown in Fig. \ref{fig:S_3D_cx_A}.

Fig. \ref{fig:S_3D_cx_A}b represents the speed of light at every
event $(t,r)$. Starting from the event $(t,r)=(0,0)$ we may thus
determine -- by numerical integration -- the past trajectory $r_{T}(t)$
of a light ray that reaches the center at the present time. That trajectory
is shown in Fig. \ref{fig:rT_A_B}, see also Fig. \ref{fig:Two-trajectories}
in the Appendix, Chapter 7. It is only from points on this trajectory
that we, at $(0,0)$, can receive information about light-emitting
stars.

The 3-D plots \ref{fig:sig_3D_A} through \ref{fig:S_3D_cx_A} exhibit
the values of the fields $Z(t,r_{T}(t))$ through $\hat{c}_{x}(t,r_{T}(t))$
along the solid lines starting at $(t,r)=(0,0)$and ending on the
surface of the sphere.

The initial values that start from $(\hat{t},\hat{r})=(-0.677,0)$
are also shown as solid lines in the 3-D plots 5 through 8. Note that
those lines in the $(\hat{\sigma},\hat{t},\hat{r})$-plot and in the
$(\hat{m},\hat{t},\hat{r})$-plot represent our trial functions (\ref{eq:AB_rho_m})
for the ``good'' parameter choice A.

The radius $R(t)$ and the trajectory $r_{T}(t)$ intersect for choice
A at the event $(\hat{t}_{0},\hat{r}_{0})=(-0.612,0.489)$. That event
is the farthest in time and space which we can observe. It occurred
at a time when the size of the universe was roughly half its present
size; indeed we have $r_{0}\approx0.5R(0)$.

Naturally the values of our fields along the trajectory are most important
for the astronomer. We denote them by $\hat{\sigma}(\hat{t})\equiv\hat{\sigma}(\hat{t},\hat{r}_{T}(\hat{t}))$
etc., if represented as functions of $\hat{t}$ with $\hat{t}_{0}\le\hat{t}\le0$
, or as $\hat{\sigma}(\hat{r})$ etc., if represented as functions
$r$ with $0\le r\le r_{0}$.

Fig. \ref{fig:sig_3D_A}b shows $\hat{\sigma}(\hat{r})$; it exhibits
a non-monotone character due to the competing facts that $\sigma$
was larger in the past in an expanding universe but has to drop to
zero at the surface of the universe. The starting value $\hat{\sigma}\approx4$
means that the central density $\sigma(0,r_{T}(0))$ at the present
time is four times bigger than the mean density at that time, see
(\ref{eq:nondim-var})\textsubscript{4}.

\begin{figure}
\includegraphics[scale=0.8]{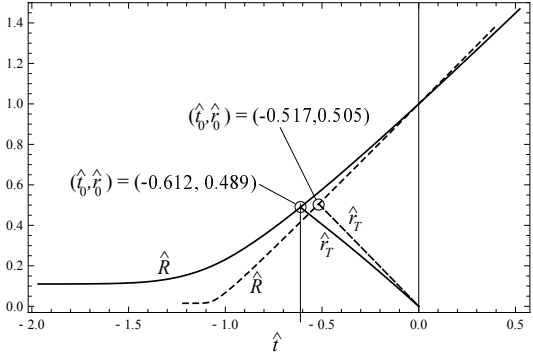}\caption{\label{fig:rT_A_B}$\hat{r}_{T}(t)$ is the light trajectory. Parameter
choice A (solid) and B (dashed). Note that $\hat{r}_{T}(t)$ is slightly
concave, see also Fig. \ref{fig:Two-trajectories} in the Appendix,
Chapter 7.}
\end{figure}

Note that the slope of the light trajectory in Fig. \ref{fig:rT_A_B}
is the speed of light $c_{x}$ at the event $(t,r_{T}(t))$. It is
everywhere slightly smaller than $c$, but it grows as the central
event $(0,0)$ is approached.

Although the 3-D plots of Figs. \ref{fig:sig_3D_A} through \ref{fig:S_3D_cx_A}
represent the solution of the Einstein equations, it is best to postpone
a discussion of its qualitative and quantitative features. Indeed,
the solution is represented here in terms of Schwarzschild coordinates
$(t,r)$. These coordinates have allowed us a fairly easy solution,
but they are bad for interpretation in terms of intuitive notions
of space and time, particularly time. A case in point is the asymptotic
behavior of $R(t)$ at small times, where $R$ approaches the Schwarzschild
radius asymptotically, whereas we expect an increasingly rapid decrease
of $R$ toward $Q$ under the action of the gravitation%
\footnote{Note that the graph $R(t)$ may be read forwards or backwards. In
the forward mode, -- from left to right -- it represents expansion
and in the backward mode it represents contraction.%
}. According to our solution, however, the derivative $dt/dR$ is singular
for $R\rightarrow Q$ and that forces us to deliberate about the rates
of clocks in a gravitational field as compared to their rates in a
Lorentz frame. We postpone this topic because at the present stage
we do not yet need it; see however Chapter 5.

What we do discuss next is the question whether our solution, -- albeit
in Schwarzschild coordinates --, does allow us to derive the observed
$\mu(z)$-relation of Fig. \ref{fig:mue-z-A} as we have anticipated.
Since that relation does not explicitly contain $r$ and $t$, it
is valid irrespective of our choice of space-time coordinates.

\subsection{Cosmological conundrum}

\begin{figure}
\includegraphics[scale=0.8]{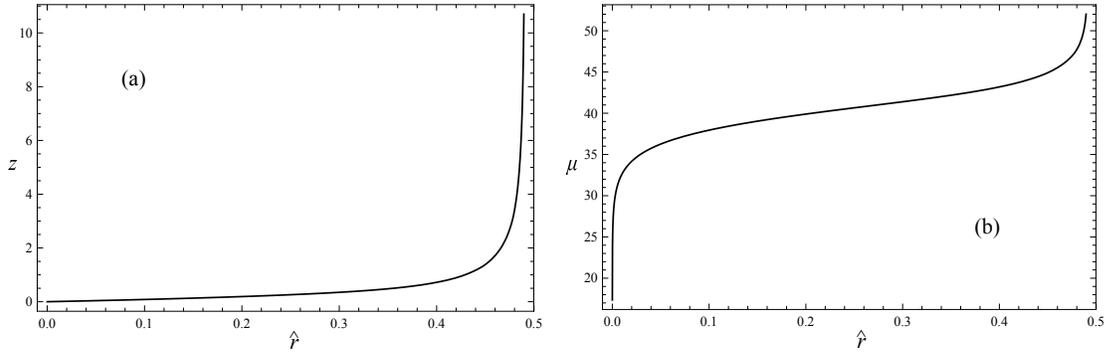}\caption{\label{fig:z_and_mue_along_rT}(a): $z(r)$. (b): and $\mu(r)$. Along
the trajectory. Parameter choice A. }
\end{figure}

In the Appendix, Chapter 7, we have derived expressions for $z(t)$
and $\mu(t)$, see also (\ref{eq:z_und_mue_von_t}). And now we have
exhibited all functions on the right-hand-sides of these, -- namely
$r_{T}(t)$, $V(t,r_{T}(t))$, $c_{x}(t,r_{T}(t))$, $S(t,r_{T}(t))$
--, so that we may plot $z(t)$ and $\mu(t)$, or $z(r)$ and $\mu(r)$.
Those plots are shown in Fig. \ref{fig:z_and_mue_along_rT} for choice
A of the parameters. Elimination of $t$ or $r$ provides the graph
$\mu(z)$ of Fig. \ref{fig:mue-z-A} which fits the observed data
well. This, of course, is a foregone conclusion since choice A represents
a ``good'' choice in the sense of Table \ref{tab:Two-good-choices}.
And the criterion for ``goodness'' was a good fit of calculated
values of the $\mu(z)$-curve with the measured data.

\begin{figure}
\includegraphics{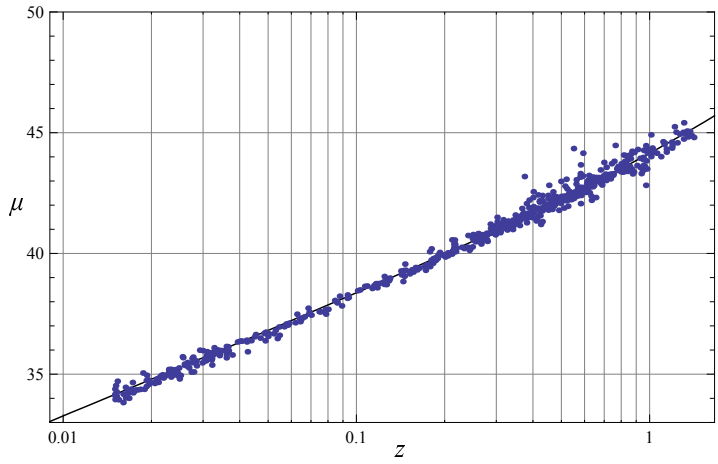}\caption{\label{fig:mue_z_B}The $\mu(z)$ -plot for the parameter choice B
of the universe.}
\end{figure}

Choice B is another good choice and accordingly the $\mu(z)$ plot
for choice B, shown in Fig. \ref{fig:mue_z_B}, is again good. In
fact, to the naked eye there is no difference between Figs. \ref{fig:mue-z-A}
and \ref{fig:mue_z_B}.

We conclude that the $\mu(z)$ results of Figs. \ref{fig:mue-z-A}
and \ref{fig:mue_z_B} do not call for a dark-energy-hypothesis nor
do they require an accelerated expansion as we shall see in Chapter
5. That is satisfactory! But there is also non-uniqueness: Although
the parameter choices A and B predict the \textit{same} Hubble diagram,
they do also predict \textit{different} density functions $\sigma(t)$
or $\sigma(r)$ along the trajectories as is illustrated in Fig. \ref{fig:sig(rT,t)_A_B}.
Other functions, like $m(r)$ or $Z(r)$ will differ as well between
the choices A and B. This means that we need more observations; observations
other than those of $\mu$ and $z$.

\begin{figure}
\includegraphics{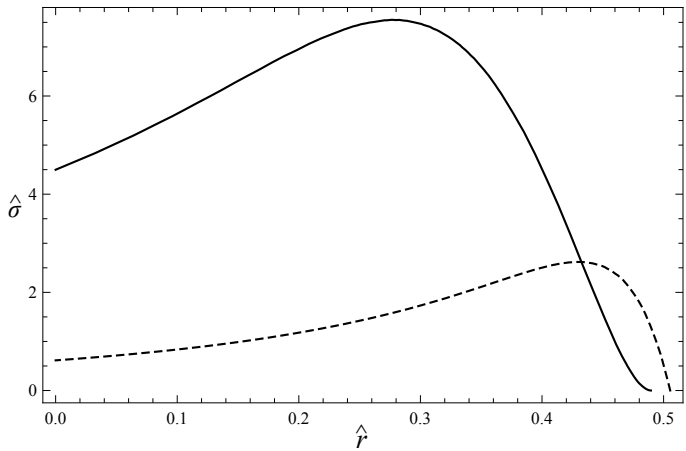}\caption{\label{fig:sig(rT,t)_A_B}Density distribution along the trajectory.
Choice A (solid), choice B (dashed).}
\end{figure}

For instance, suppose that astronomers had measured $\sigma(r)$ along
the trajectory. We should then have to perform our inverse adjustment
of initial conditions not only for the observed Hubble function $\mu(z)$
as the target function but also for the observed values $\sigma(r)$.
However, $\sigma(r)$-observations do not seem to be available. What
comes closest to them are galaxy counts.

Indeed, among the cosmological observations other than the $\mu(z)$-plot
of Fig. \ref{fig:mue-z-A} galaxy redshift surveys \cite{Colless2001}
have furnished a reliable histogram of the fraction of $250.000$
galaxies over the redshift of the light emitted by them. The histogram
has gelled into a simple smooth analytic approximation curve of the
form (see \cite{Colless2001} p. 1059ff)
\begin{equation}
fraction=z^{2}\exp(-49.3z^{1.55}).\label{eq:fraction-v-z}
\end{equation}
This formula may be converted into a \textquotedbl{}smeared out\textquotedbl{}
galaxy density, and there is the temptation to consider this galaxy
density as proportional to the matter density $\sigma$. However,
we are not certain whether far-away galaxies have the same mass as
close ones. Therefore we have not incorporated galaxy counts -- reliable
as they may be -- into our adaptive inverse scheme. That remains to
be done in the future after more deliberation.

Moreover, it can hardly be expected that an eventual observation of
$\sigma(r)$ would suffice to identify unique initial conditions.
More observational evidence is likely to be needed and the question
is: Which additional observations are feasible and what solutions
do they entail? That question represents the cosmological conundrum.

\section{Interpretation of results}

\subsection{Radius of the universe as a function of the proper time $\boldsymbol\tau \textbf{=} \boldsymbol\theta  \textbf{(} \textbf{\textit{t}} \textbf{,0)}$
at the center $\textbf{\textit{r}} \textbf{=0}$.}

As was mentioned before, -- toward the end of Section 4.5 --, the
use of Schwarzschild coordinates for the solution of the Einstein
equations has made it difficult, or impossible, to interpret the results
in terms of our intuitive notion of time. We believe that it is the
proper time
\[
\tau=\theta(t,0)
\]
 of the observer at the center of the universe that comes closest
to our intuitive notion of time and that
\begin{equation}
V_{\tau}(t,r)=\left(\frac{\partial r}{\partial\tau}\right)_{\varrho}=\frac{V(t,r)}{\frac{d\tau}{dt}\left(t\right)}\label{eq:V_tau_eqn}
\end{equation}
corresponds to our intuitive notion of velocity. Therefore we proceed
to replace $t$ by $\tau(t)$.

\begin{figure}
\includegraphics{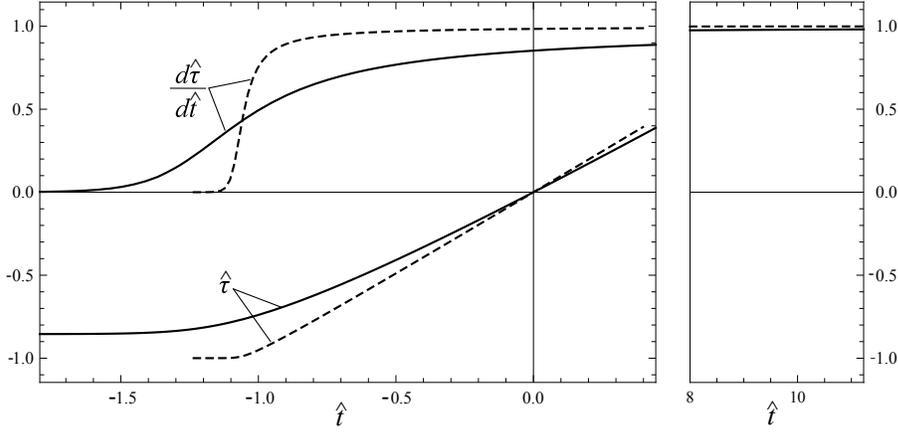}\caption{\label{fig:The-proper-time}The proper time as a function of $t$.
Parameter choice A (solid) and B(dashed)}
\end{figure}

From (\ref{eq:d_tau}) and the knowledge of $V(t,r)$, $Z(t,r)$,
and $S(t,r)$, see Figs. \ref{fig:V_3D_A},\ref{fig:Z_3D_A},\ref{fig:S_3D_cx_A},
we obtain $\frac{d\theta}{dt}(t,r)$ and, in particular, the graph
$\frac{d\tau}{dt}(t)=\frac{d\theta}{dt}(t,0)$ of Fig. \ref{fig:The-proper-time}.
Integration over $t$ gives $\tau(t)$ to within a constant which
we identify by requiring that $\tau=0$ holds for $t=0$. The resulting
graph $\tau(t)$ is also shown in Fig. \ref{fig:The-proper-time}.
Inspection shows that for large values of $t$ the proper time $\tau$
becomes asymptotically equal to $t$, while for small values of $t$
the proper time becomes independent of $t$.

\begin{figure}
\includegraphics[scale=0.8]{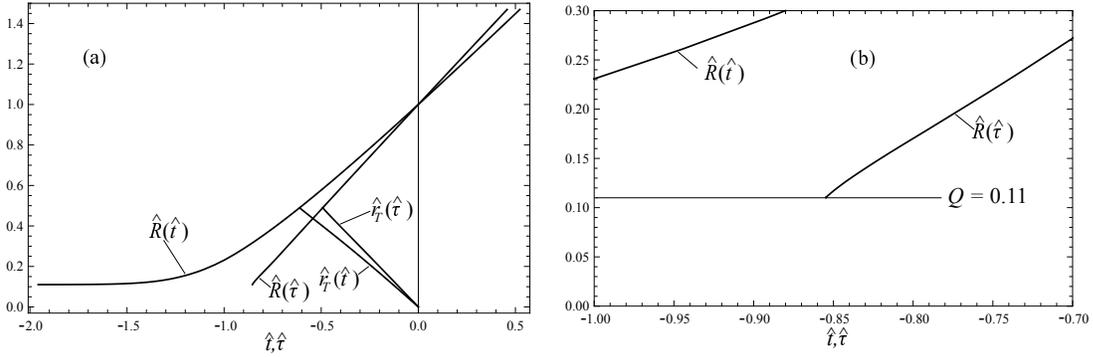}\caption{\label{fig:Rvt_and_Rvtau_A}Parameter choice A. Radii $R(t)$ and
$R(\tau)$. Also trajectories $r_{T}(t)$ and $r_{T}(\tau)$.}
\end{figure}

We conclude that sufficiently far in the past the tiniest increase
of $\tau$ requires a large increase in $t$, or else: the rate of
Schwarzschild coordinate clocks, which measure $t$, is much larger
than the rate of a clock at rest in a Lorentz frame at the center.
It is for this reason that the radius $R(t)$ in Figs. \ref{fig:rT_A_B}
and \ref{fig:Rvt_and_Rvtau_A} does not seem to change at all with
time $t$ as the radius approaches the Schwarzschild radius $Q$.

\begin{figure}
\includegraphics{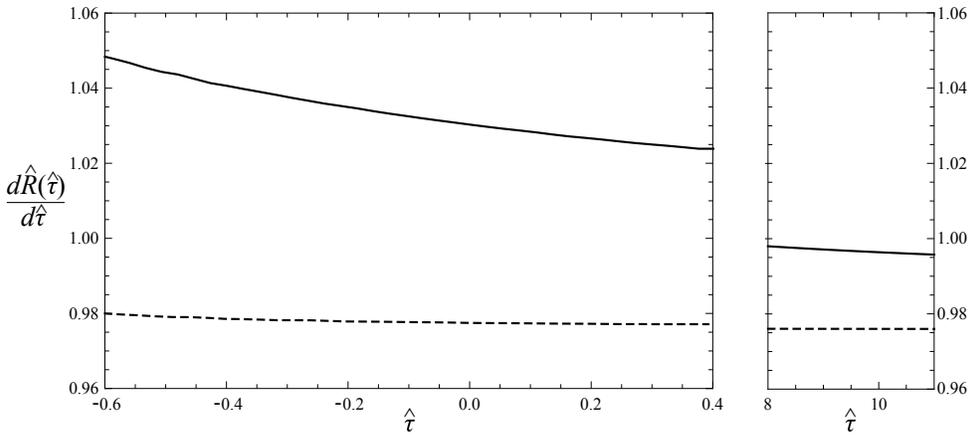}\caption{\label{fig:Surface-velocity}Surface velocity $dR(\tau)/d\tau$ in
the range $-0.6<\tau<0.4$ and $8<\tau<11$. Choice A (solid), choice
B (dashed).}
\end{figure}

$R(\tau)$ on the other hand -- obtained by elimination of $t$ between
$R(t)$ and $\tau=\theta(t,0)$ -- seems to grow linearly starting
with $R=Q$, see Fig. \ref{fig:Rvt_and_Rvtau_A}a which shows both
functions $R(t)$ and $R(\tau)$. Closer inspection, however, reveals
that $R(\tau)$ is a slightly concave function so that the observer
at $r=0$ sees a decelerating expansion in terms of \textit{his} time
$\tau$. Fig. \ref{fig:Rvt_and_Rvtau_A}b demonstrates this fact by
focussing the attention on the neighborhood of the point where $R(\tau)$
emerges from the Schwarzschild radius; that happens at approximately
$\tau=-0.85$. The fact is further illustrated in Fig. \ref{fig:Surface-velocity}
where the velocity of expansion $dR(\tau)/d\tau$ is plotted. The
figure shows that the velocity is a convex function and has not even
developed into a straight line at $\tau=0$, i.e. at the present time.
For large values of $\tau$ the surface velocity tends to a constant
value $<1$. For choice B this is obvious from inspection of Fig.
\ref{fig:Surface-velocity}; for choice A we need to pursue the graph
far to right in order to see it drop below the value 1.

\begin{figure}
\includegraphics[scale=0.8]{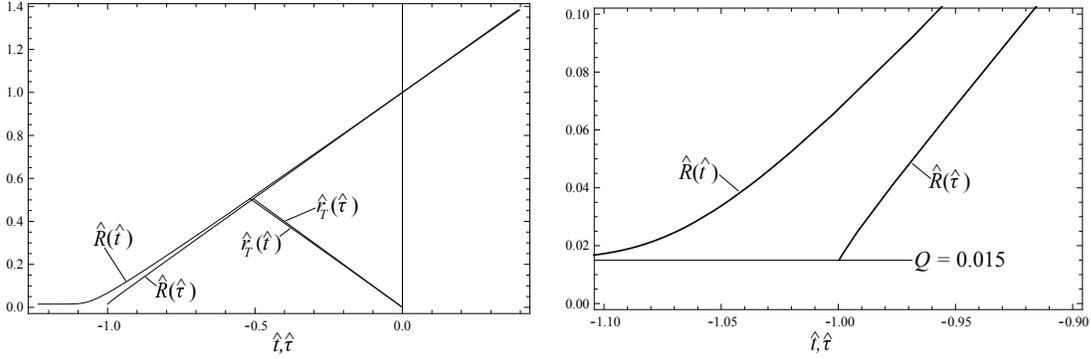}\caption{\label{fig:Rvt_and_Rvtau_B}Parameter choice B. Radii $R(t)$ and
$R(\tau)$. Also trajectories $r_{T}(t)$ and $r_{T}(\tau)$.}
\end{figure}

The plots of Fig. \ref{fig:Rvt_and_Rvtau_B} correspond to those of
Fig. \ref{fig:Rvt_and_Rvtau_A} except that they are calculated for
the parameter choice B. For both choices, A and B, we see the same
qualitative behaviour of the $R(\tau)$-curve. The concavity in that
curve -- as exhibited in Fig. \ref{fig:Rvt_and_Rvtau_B}b -- is a
little more pronounced for choice B than for choice A. From both figures
we conclude that the expansion of the universe is decelerating, since
$\hat{R}(\hat{\tau})$ is concave.

\phantom{ }

\footnotesize The significant difference between $R(t)$ and $R(\tau)$
in Figs. \ref{fig:Rvt_and_Rvtau_A} and \ref{fig:Rvt_and_Rvtau_B}
-- one convex and the other concave -- is reminiscent of the case
of a mass which drops into a black hole. That case is routinely treated
in books on general relativity, e.g. see Shapiro and Teukolsky \cite{Shapiro1983}.
There too the distance $h(t)$ of the mass from the black hole tends
to a finite value as the Schwarzschild radius of the mass is approached,
while $h(\tau)$, where $\tau$ is the proper time of the mass, drops
precipitously toward zero.\normalsize

\phantom{ }

\textit{Our} solution does not provide graphs for $R(\tau)<Q$ simply
because we do not get values for $R(t)<Q$. What we do obtain is the
proper time when the expanding sphere emerges from the Schwarzschild
radius. That time may be read off from the figures as the abscissae
of the end points of the $R(\tau)$-curve and maybe that time is appropriately
called the \textquotedbl{}time of emergence.\textquotedbl{} We know
nothing about the period of time before that age, which was where
the \textquotedbl{}big bang\textquotedbl{} happened -- if it happened
-- and where the expansion began. Our cosmological model is not able
to provide information about such phenomena, at least not in the present
form.

\subsection{Trajectories}

The Figs. \ref{fig:Rvt_and_Rvtau_A} and \ref{fig:Rvt_and_Rvtau_B}
also show graphs of the trajectories $r_{T}(t)$ and of $r_{T}(\tau)$;
the latter results by elimination of $t$ between $r_{T}(t)$ and
$\tau(t)$. It is only from events on these curves that we may receive
information by light about the past. In analogy to Fig. \ref{fig:rT_A_B}
we denote the coordinates of the points of intersection of the graphs
$r_{T}(\tau)$ and $R(\tau)$ by $(\tau_{0},r_{0})$. We cannot observe
what happened before $\tau_{0}$ and at a greater distance than $r_{0}$.
Therefore $\tau_{0}$ may be called a \textquotedbl{}\textit{look
back time}.\textquotedbl{} Its value is given in Table \ref{tab:Some-results-for}
along with other interesting properties of our models.

While the trajectory $r_{T}(t)$ is very slightly concave, the graph
$r_{T}(\tau)$ is straight and its slope is equal to 1 which, in dimensional
terms, means that -- of course -- the light moving along $r_{T}(\tau)$
has the speed $c$ of light in the Lorentz frame at the center.

\subsection{Some results in tabular form}

We summarize some of our results in Table \ref{tab:Some-results-for}.
We recall that $R(0)$ was also listed in Table \ref{tab:Two-good-choices}
and that it was determined by adjusting our parameters to give a good
agreement between theory and observation of the $\mu(z)$-curve. Given
$R(0)$ the total mass follows from the definition (\ref{eq:Q_eqn})
of the Schwarzschild radius.

\begin{table}[H]
\caption{\label{tab:Some-results-for}Some results for the parameter choice
A and B}
\begin{tabular}{|c|c|c|c|c|c|c|}
\hline 
 & $R(0)$ & $M$ & Time of emergence & $\tau_{0}$ & $r_{0}$ & $H_{0}$\tabularnewline
\hline 
\hline 
A & 10.79 Gly & $1.514\cdot10^{52}$kg & $\approx-0.85\cong-9.22$Gy & $\cong-5.311$Gy & $\cong5.281$Gly & $0.22\cdot10^{-17}$s\textsuperscript{-1}\tabularnewline
\hline 
B & 14.27 Gly & $0.273\cdot10^{52}$kg & $\approx-1.00\cong-14.26$Gy & $\cong-7.212$Gy & $\cong7.210$Gly & $0.22\cdot10^{-17}$s\textsuperscript{-1}\tabularnewline
\hline 
\end{tabular}
\end{table}

The data given in the table for choices A and B differ widely. And,
what is more, they differ considerably from the values provided by
the currently popular FRW cosmology, where, for instance, the age
of the universe is confidently given as $-13.7$Gy. Well, we can only
say that that value is heavily dependent on the Robertson-Walker model
along with the concept of a hypothetical dark energy which, moreover,
makes up 70\% (!) of all mass in the universe. We do not consider
that notion convincing.

\subsection{Hubble\textasciiacute{}s law}

\begin{figure}
\includegraphics{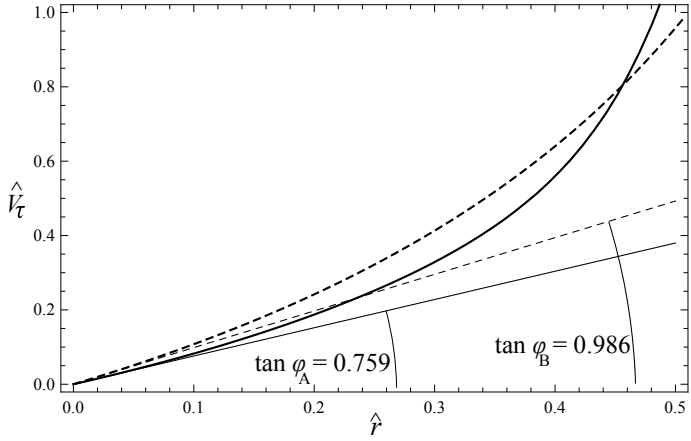}\caption{\label{fig:Vtau(r)_A_B}Velocity as function of distance along the
trajectory. Choice A (solid), B (dashed). The slope of the tangent
may be identified with the Hubble constant.}
\end{figure}

The very fact that the function $t=t(\tau)$, according to Fig. \ref{fig:The-proper-time},
has a singularity for $\tau\approx-0.85$ disqualifies $t$ as an
appropriate measure for time. That fact makes it impossible to interpret
$V(t,r)=\left(\frac{\partial r}{\partial t}\right)_{\varrho}(t,r)$
as a reasonable measurable velocity. Therefore we ought to reformulate
the calculated field $V(t,r)$ in terms of the proper time of the
observer in the center. As in (\ref{eq:V_tau_eqn}), we consider
\[
V_{\tau}(t,r)=\frac{V(t,r)}{\frac{d\tau}{dt}(t)}
\]
as the proper measure of velocity. This is a function on the $(t,r$)-plane.
From it the velocity $V_{\tau}(t,r_{T}(t))$ along the trajectory,
i.e. the velocity of a star which has sent out light at a time $t<0$,
may be calculated. Hence follows the velocity $V_{\tau}(r)$ of the
star at the distance $r$ along the trajectory by elimination of $t$
between $V_{\tau}(t,r_{T}(t))$ and $r_{T}(t)$ . It is shown in Fig.
\ref{fig:Vtau(r)_A_B} and we see that it increases with an increasing
gradient as $r$ approaches the surface of the sphere so that on the
surface -- where $r=r_{0}$ -- the velocity $V_{\tau}(r)$ is close
to 1. This means that the surface moves away from us roughly with
the speed of light.

Close to the center, where Hubble \cite{Hubble1929} made his observations,
the convex curve $V_{\tau}(r)$ may be approximated by its tangent
whose slope is the Hubble constant $\hat{H}_{0}=0.759$ for choice
A, see Fig \ref{fig:Vtau(r)_A_B}. In dimensional form this value
is thus given by $H_{0}=\hat{H}_{0}\frac{c}{R(0)}=0.22\cdot10^{-17}\textrm{s}^{-1}$;
it is listed in Table \ref{tab:Some-results-for}. This value agrees
well with the value $H_{0}=0.226\cdot10^{-17}\textrm{s}^{-1}$ which
is most often given in the astronomical literature, e.g. see \cite{Weinberg2008}.

For choice B the initial slope of the $V_{\tau}(r)$ curve is given
by $\hat{H}_{0}=0.986$ which again implies $H_{0}=0.22\cdot10^{-17}\textrm{s}^{-1}$
in dimensional form, see Table \ref{tab:Some-results-for}. We must
not be surprised that the choices A and B agree as to the value of
the Hubble constant despite their disagreement about all other values
in Table \ref{tab:Some-results-for}. Because indeed, the models do
agree on the $\mu(z$)-relation, see Figs. \ref{fig:mue-z-A} and
\ref{fig:mue_z_B} and Hubble\'{ }s law is a corollary of that relation.
We proceed to describe this point briefly.\\

\footnotesize Generally we have, by (\ref{eq:mue_eqn})\textsubscript{1}
in the Appendix, Chapter 7 
\[
d_{L}=10\,\textrm{pc}\,10^{\frac{\mu(z)}{5}},
\]
where $d_{L}$ is the luminosity distance defined in (\ref{eq:d_Luminosity}).
For small values of $z$ we have from the observed data, see Figs.
\ref{fig:mue-z-A} and \ref{fig:mue_z_B}
\[
\mu(z)\approx33.38+5\log_{10}\frac{z}{0.01}.
\]
Also, by (\ref{eq:d_Luminosity})\textsubscript{1} and (\ref{eq:z_eqn})
we have for small $z$
\[
d_{L}\approx r\;\textrm{and}\; z\approx\hat{V}_{\tau}.
\]
Hence follows for the non-dimensional Hubble constant
\[
\hat{H}_{0}=\frac{\hat{V}_{\tau}}{\hat{r}}=\frac{R(0)}{10\,\textrm{pc}}\frac{1}{10^{\frac{43.38}{5}}}.
\]
For the dimensional Hubble constant $H{}_{0}=\frac{c}{R(0)}\hat{H}_{0}$
the choice-dependent factor $R(0)$ cancels and thus $H_{0}$ is the
same for choices A and B: Hubble's observations for small $z$ have
no independent status vis-a-vis the $\mu(z)$ observations.\normalsize

\subsection{Accelerating and decelerating contributions to the expansion of the
universe}

We have argued that the Schwarzschild coordinate time $t$ is not
a good measure of time and that the proper time $\tau=\theta(t,0)$
in the center of the sphere is better. The difference is not trivial;
indeed $R(t)$ exhibits an accelerated growth while $R(\tau)$ exhibits
a decelerated growth, see Section 5.1. To be sure, according to Figs.
\ref{fig:Rvt_and_Rvtau_A} and \ref{fig:Rvt_and_Rvtau_B} both modes
of growth eventually settle into a uniform expansion with a constant
velocity close to that of light, see Fig. \ref{fig:Surface-velocity}.
This is strange in itself, quite apart from the question whether the
radius in the past has been accelerated or decelerated toward that
uniform value. Indeed, do we not expect -- as more or less classical
physicists -- that there should be a continued deceleration under
the effect of the gravitational attraction? In answer to this question
let us discuss the role of gravity in our model; that role has so
far been disguised by the unfamiliar form of the equations and by
the fog raised through the numerical solution of our inverse problem.

We call the attention to (\ref{eq:four_field_eqns})\textsubscript{4}
which, after a little rearrangement and the insertion of dimensionless
quantities reads
\begin{equation}
\frac{d\hat{V}}{d\hat{t}}=-\frac{1}{2}Q\frac{\hat{m}}{\hat{r}^{2}}+Q\frac{\hat{m}}{\hat{r}^{2}}\hat{V}^{2}\frac{1}{1-Q\frac{\hat{m}}{\hat{r}}}-\frac{1}{2}Q\frac{\hat{m}}{\hat{r}^{2}}(Z-1)+\frac{1}{2}\frac{\hat{V}}{Z}\frac{dZ}{d\hat{t}}\label{eq:V_eqn_in_t}
\end{equation}
in terms of the Schwarzschild time and%
\footnote{The hats for non-dimensional quantities are now dropped. Besides the
tilde denote functions of $\tau$ as the time variable.%
}
\begin{equation}
\frac{d\tilde{V}_{\tau}}{d\tau}=-\frac{1}{2}Q\frac{\tilde{m}}{r^{2}}+Q\frac{\tilde{m}}{r^{2}}\tilde{V}_{\tau}^{2}\frac{1}{1-Q\frac{\tilde{m}}{r}}-\frac{1}{2}Q\frac{\tilde{m}}{r^{2}}\left(\frac{\tilde{Z}(\tau,r)}{\tilde{Z}(\tau,0)}-1\right)+\frac{1}{2}\tilde{V}_{\tau}\left(\frac{1}{\tilde{Z}(\tau,r)}\frac{d\tilde{Z}(\tau,r)}{d\tau}-\frac{1}{\tilde{Z}(\tau,0)}\frac{d\tilde{Z}(\tau,0)}{d\tau}\right)\label{eq:V_eqn_in_tau}
\end{equation}
in terms of the proper time $\tau(t)$.

It is clear that those equations must be identities when we insert
the solution exhibited in Chapter 4, because the solution has been
derived from them. However, (\ref{eq:V_eqn_in_t}) and (\ref{eq:V_eqn_in_tau})
may be viewed as expressions for the accelerations $dV/dt$ or $d\tilde{V}_{\tau}/d\tau$.
We concentrate on $d\tilde{V}_{\tau}/d\tau$, because that is the
acceleration as seen by the observer in the center. The first term
on the right-hand-side is the classical gravitational acceleration.
It is clearly attractive because of the minus sign. But it is not
alone!

\begin{figure}
\includegraphics{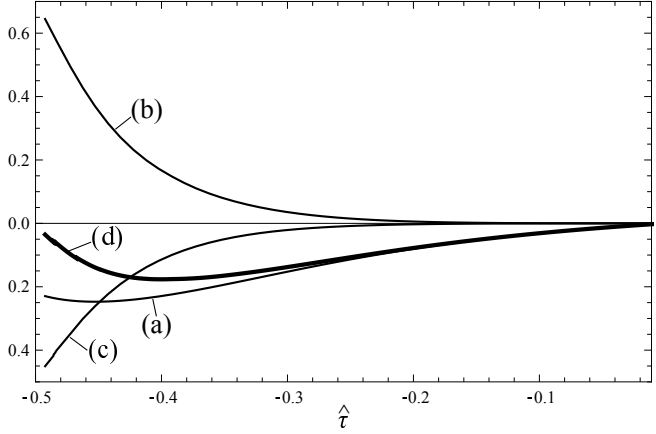}\caption{\label{fig:rhs_of_V_eqn}(a) Newtonian deceleration. (b) Repulsive
acceleration. (c) Terms with $Z$. (d) Overall deceleration along
the trajectory.}
\end{figure}

Fig. \ref{fig:rhs_of_V_eqn} shows plots of the four contributions
on the right hand side of (\ref{eq:V_eqn_in_tau}) along the trajectory
$r_{T}(t)$: In (a) we see the attractive classical (Newtonian) contribution,
while (b) shows the repulsive second term which grows with growing
velocity as the surface is approached. In (c) we have plotted the
third and fourth terms, -- the terms with $Z$ and its derivatives;
those are negative and therefore attractive. Finally the fat graph
(d) exhibits the entire right-hand-side of (\ref{eq:V_eqn_in_tau}).
We see that the classical attraction is diminished by the other terms
throughout the whole spread of the universe but that it remains attractive,
albeit weakly near the surface. We conclude that there is no accelerated
expansion anywhere. Therefore there is no need to introduce a hypothetical
dark energy to create the accelerated expansion.

Also we now understand why it is that the surface of the universe
moves with a nearly constant speed for large times despite of what
we are accustomed to consider as the gravitational pull. Indeed, the
gravitational pull is quite small on the surface.

The parameters for which the graphs of Fig. \ref{fig:rhs_of_V_eqn}
are drawn, are those of parameter choice A. For choice B we obtain
smaller values -- obviously reflecting the smaller mass $M$ of choice
B -- and the final value at $\hat{\tau}=-0.5$ remains negative: Apparently
the deceleration has not come to an end yet at the surface of the
sphere.

The various accelerations of Fig. \ref{fig:rhs_of_V_eqn} are all
calculated for the present trajectory, i.e. the trajectory that passes
through the event $(0,0)$. Surely the plots will look different for
trajectories at earlier times. And it is even conceivable that for
such earlier times the graph (d) may exhibit an overall acceleration
instead of an overall deceleration. In that respect it seems worthy
of note that for real early trajectories $Z=0$ holds and that the
accelerating term (b) in Fig. \ref{fig:rhs_of_V_eqn} becomes singular,
see Fig. \ref{fig:Z_3D_A}a and eqn. (\ref{eq:V_eqn_in_tau}). That
aspect, which touches on the notion of an inflationary past period
of the universe, will be the subject of a subsequent study, if the
present one fares well.

In the present study the main conclusion is that there is no overall
acceleration along the present trajectory and therefore there is no
need for a hypothetical dark energy.

\section{Remark on the cosmological principle}

We are fully aware of the fact that our model of a sherical universe
floating in infinite empty space violates the \textit{cosmological
principle} according to which we, the observers should not occupy
a privileged place in the cosmos. In our model clearly the center
of the sphere is privileged and yet that site is assumed to be the
place of the observer.

In early cosmology the problem became urgent with Hubble's observation
that the galaxies in our neighbourhood move away from us with a speed
proportional to their distance. The question was: Why are we thus
privileged? That question lost some of its urgency when it was recognized
that in an infinitely extended homogeneously dense matter distribution
with an expansive velocity proportional to the distance from \textit{one}
point the matter is in fact moving away from \textit{all} points in
the same manner. So, in such a universe Hubble's observation does
not put us in a privileged spot.

\begin{figure}
\includegraphics[scale=0.8]{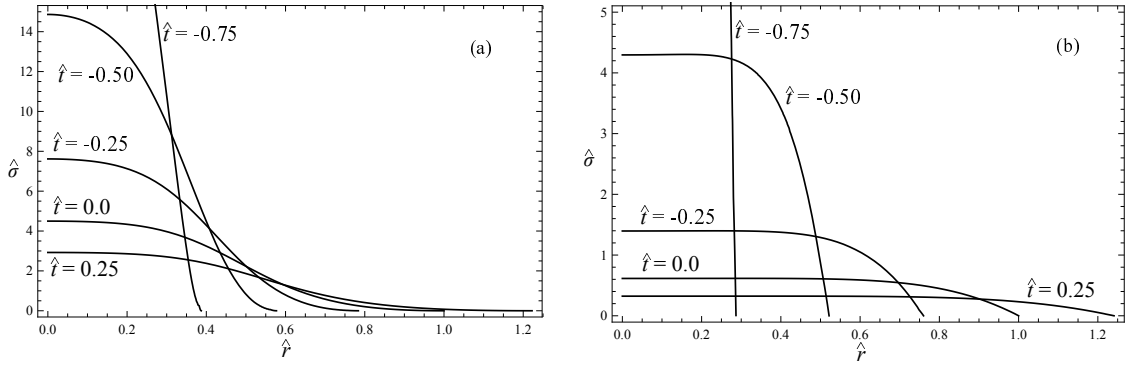}\caption{\label{fig:sig_A_B}$\sigma(t,r)$ for $0\leq r\leq R(t)$ at various
times. (a): Choice A. (b): Choice B.}
\end{figure}

\begin{figure}
\includegraphics[scale=0.8]{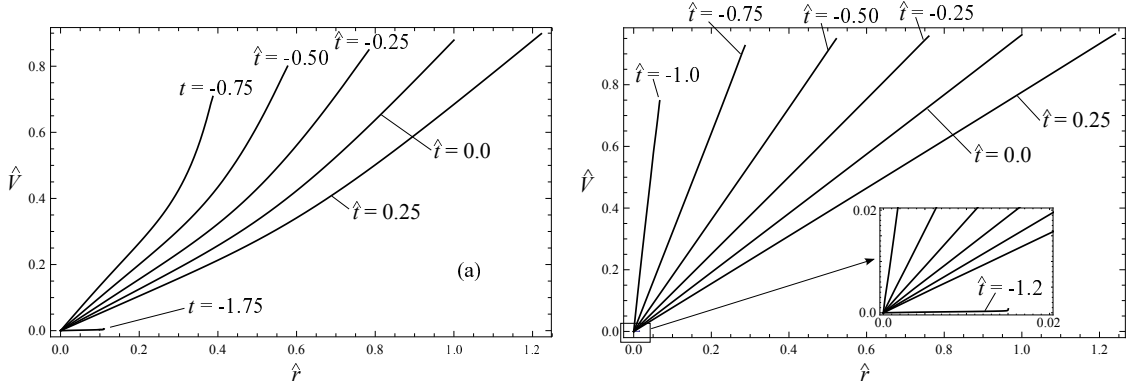}\caption{\label{fig:Vt_A_B}$V(t,r)$ for $0\leq r\leq R(t)$ at various times.
(a): Choice A. (b): Choice B.}
\end{figure}

Of course, if the universe is spherical with a finite radius and not
homogeneously filled with matter -- as in our case -- the argument
does not hold, or at least it does not hold exactly and everywhere.
We may conjecture, however, that the argument \textit{does} hold approximately
in an inner sphere around the center where the density is nearly constant
and where the matter moves away from the center in the Hubble way.
Fig. \ref{fig:sig_A_B} and \ref{fig:Vt_A_B} show that there \textit{is}
such an inner sphere, well pronounced for choice B and somewhat less
well pronounced for choice A.

Clearly that argument needs strengthening. It is offered here loosely
so as to anticipate the objection that our model of the universe places
the observer in the central \textit{point} of the universe. We believe
that the model is still good, if the observer is placed anywhere within
the inner sphere of near-homogeneous density.

\section{Appendix: Redshift of Type Ia supernovae and luminosities.}

\subsection{Scope}

The study of redshifts requires an investigation of the Doppler shift
and aberration and of the gravitational frequency shift. We discuss
these phenomena in the subsequent Sections 7.2 through 7.4 with (\ref{eq:z_eqn})
and Fig. \ref{fig:z_and_mue_along_rT}a as the result. Section 7.5
is given to a discussion of the apparent luminosity with (\ref{eq:mue_eqn})\textsubscript{2}
and Fig. \ref{fig:z_and_mue_along_rT}b as the result. Both results
(\ref{eq:z_eqn}) and (\ref{eq:mue_eqn})\textsubscript{2} have been
anticipated in Sections 4.1 and 4.6.

The formulae for redshifts and luminosities differ between different
cosmological models, because the concepts of frequency and wave length
are non-trivial in relativity. Etherington \cite{Etherington1933}
has studied the problem as early as 1933. See also Ellis \cite{Ellis2007}
who reviews Etherington\textasciiacute{}s work as a \textquotedbl{}Golden
Oldie.\textquotedbl{} Still, however, there are conflicting formulae
used as relations between absolute and apparent luminosities in the
modern literature. Thus the relevant results of the Friedmann-Robertson-Walker
model are useless for our more mundane model; there is not even a
clear-cut Doppler effect in the FRW-model, see \cite{Weinberg2008}.
Therefore we have rederived the relevant formula in this chapter and
we hope and trust that we got things right.

\subsection{Dopplershift and aberration}

In this chapter we consider light emitted by a star or galaxy moving
with the radial velocity $V$ with respect to the center of the sphere
and absorbed in the center.

We look at the light from three different frames of reference: i.)
the local and momentary Lorentz frames ${z^{\alpha}}$ and ${z_{0}^{\alpha}}$
-- with coordinates ${\theta\,,z^{1},z^{2},z^{3}}$ and ${\theta_{0}\,,z_{0}^{1},z_{0}^{2},z_{0}^{3}}$
respectively -- which accompany the star and the observer and ii.)
the frame ${x^{\alpha}}$ -- with coordinates ${t\,,x{}^{1},x{}^{2},x{}^{3}}$
-- in which we have solved the Einstein equations. It is now appropriate
to introduce rectangular Cartesian spatial coordinates ${z^{\alpha}}$
and ${x^{\alpha}}$, since the light does not necessarily move in
the radial direction. However, we let $z^{1}$ and $x^{1}$ point
into the radial direction so that the unit vectors of propagation
of the light may be given in terms of the direction angles $\alpha$
and $\beta$ by
\begin{equation}
\underset{z}{n}^{a}=\left(\cos\underset{z}{\alpha},\sin\underset{z}{\alpha}\cos\underset{z}{\beta},\sin\underset{z}{\alpha}\sin\underset{z}{\beta}\right)\:\textrm{and}\:\underset{x}{n}^{a}=\left(\cos\underset{x}{\alpha},\sin\underset{x}{\alpha}\cos\underset{x}{\beta},\sin\underset{x}{\alpha}\sin\underset{z}{\beta}\right).
\end{equation}
Let the light locally be described as a plane wave with frequency
$\nu$ and wave length $\lambda$. The space-time wave vector $k_{\alpha}$
is then given by
\[
\underset{z}{k_{\alpha}}=2\pi\left(\frac{\underset{z}{\nu}}{c},-\frac{1}{\underset{z}{\lambda}}\underset{z}{n}^{a}\right),\:\textrm{where}\:\underset{z}{\nu}\underset{z}{\lambda}=c
\]
\begin{equation}
\underset{x}{k_{\alpha}}=2\pi\left(\frac{\underset{x}{\nu}}{c},-\frac{1}{\underset{x}{\lambda}}\underset{x}{n}^{a}\right),\:\textrm{where}\:\underset{x}{\nu}\underset{x}{\lambda}=\underset{x}{c}.\label{eq:k_alfa_x}
\end{equation}
We consider $\underset{z}{k_{\alpha}}$ as given and calculate $\underset{x}{k_{\alpha}}$
using the transformation matrix between the frames $\left\{ z^{\alpha}\right\} $
and $\left\{ x{}^{\alpha}\right\} $
\begin{equation}
\underset{x}{k_{\alpha}}=\frac{\partial z^{\beta}}{\partial x^{\alpha}}\underset{z}{k_{\beta}}.\label{eq:k_alfa_x_2}
\end{equation}
The transformation matrix may be calculated from the invariance of
the infinitesimal distance element in space-time, see (\ref{eq:ds_eqn}),
(\ref{eq:ds_eqn_eigen})
\begin{equation}
ds^{2}=-c^{2}d\theta^{2}+dz^{12}+dz^{22}+dz^{32}=Zc^{2}dt^{2}+Sdx^{12}+Sdx^{22}+Sdx^{32}.\label{eq:ds_quadrat_2}
\end{equation}
The calculation provides
\begin{equation}
\frac{\partial z^{\beta}}{\partial x^{\alpha}}=\left(\begin{array}{cccc}
\frac{\sqrt{\left|Z\right|}}{\sqrt{1-\frac{V^{2}}{\underset{x}{c}^{2}}}} & -\frac{V}{\underset{x}{c}}\frac{\sqrt{\left|Z\right|}}{\sqrt{1-\frac{V^{2}}{\underset{x}{c}^{2}}}} & 0 & 0\\
-\frac{V}{\underset{x}{c}}\frac{\sqrt{S}}{\sqrt{1-\frac{V^{2}}{\underset{x}{c}^{2}}}} & \frac{\sqrt{S}}{\sqrt{1-\frac{V^{2}}{\underset{x}{c}^{2}}}} & 0 & 0\\
0 & 0 & \sqrt{S} & 0\\
0 & 0 & 0 & \sqrt{S}
\end{array}\right).\label{eq:dz_beta_dx_alfa}
\end{equation}
As before, $V=\left(\frac{\partial r}{\partial t}\right)_{\varrho}$
is the velocity of the star which emits the light; it is non-negative
for the expanding sphere. Insertion of (\ref{eq:dz_beta_dx_alfa})
into (\ref{eq:k_alfa_x_2}) gives
\begin{equation}
\frac{\underset{x}{\nu}}{\underset{z}{\nu}}=\sqrt{\left|Z\right|}\frac{1+\frac{V}{\underset{x}{c}}\underset{z}{n}^{1}}{\sqrt{1-\frac{V^{2}}{\underset{x}{c}^{2}}}},\;\frac{\underset{x}{\nu}}{\underset{z}{\nu}}\underset{x}{n}^{1}=\sqrt{\left|Z\right|}\frac{\underset{z}{n}^{1}+\frac{V}{\underset{x}{c}}}{\sqrt{1-\frac{V^{2}}{\underset{x}{c}^{2}}}},\;\frac{\underset{x}{\nu}}{\underset{z}{\nu}}\underset{x}{n}^{2}=\sqrt{\left|Z\right|}\underset{z}{n}^{2},\;\frac{\underset{x}{\nu}}{\underset{z}{\nu}}\underset{x}{n}^{3}=\sqrt{\left|Z\right|}\underset{z}{n}^{3}.\label{eq:nue_x/nue_z}
\end{equation}
Hence follows in terms of the direction angles $\alpha$ and $\beta$
of the light
\begin{equation}
\frac{\underset{x}{\nu}}{\underset{z}{\nu}}=\sqrt{\left|Z\right|}\frac{1+\frac{V}{\underset{x}{c}}\cos\underset{z}{\alpha}}{\sqrt{1-\frac{V^{2}}{\underset{x}{c}^{2}}}},\:\cos\underset{x}{\alpha}=\frac{\cos\underset{z}{\alpha}+\frac{V}{\underset{x}{c}}}{1+\frac{V}{\underset{x}{c}}\cos\underset{z}{\alpha}},\:\underset{x}{\beta}=\underset{z}{\beta}.\label{eq:doppler_and_aberration_eqns}
\end{equation}
These are the equations of Doppler shift and aberration, so called
in analogy to similar effects in acoustics. The aberration equations
(\ref{eq:doppler_and_aberration_eqns})\textsubscript{2,3} imply
for an element $d\Omega=\sin\alpha d\alpha d\beta$ covered by light
rays of neighbouring directions
\begin{equation}
d\underset{x}{\Omega}=\frac{1}{1+\frac{V}{\underset{x}{c}}\cos\underset{z}{\alpha}}\left(1-\frac{\cos\underset{z}{\alpha}+\frac{V}{\underset{x}{c}}}{1+\frac{V}{\underset{x}{c}}\cos\underset{z}{\alpha}}\frac{V}{\underset{x}{c}}\right)d\underset{z}{\Omega}.\label{eq:d_Omega_x}
\end{equation}
In particular, when the light moves radially inwards, i.e. for $\underset{z}{\alpha}=\pi$,
we obtain
\begin{equation}
\frac{\underset{x}{\nu}}{\underset{z}{\nu}}=\sqrt{\left|Z\right|}\sqrt{\frac{1-\frac{V}{\underset{x}{c}}}{1+\frac{V}{\underset{x}{c}}}}\:\textrm{and}\: d\underset{x}{\Omega}=\frac{1+\frac{V}{\underset{x}{c}}}{1-\frac{V}{\underset{x}{c}}}d\underset{z}{\Omega}\label{eq:nue_x_over_Nue_z_and_d_Omega_x}
\end{equation}
so that the Doppler shift is a red shift in this case and the solid
angle element is bigger in frame $\left\{ x{}^{\alpha}\right\} $
than in the Lorentz frame $\left\{ z^{\alpha}\right\} $.

\subsection{Gravitational shift}

\begin{figure}
\includegraphics{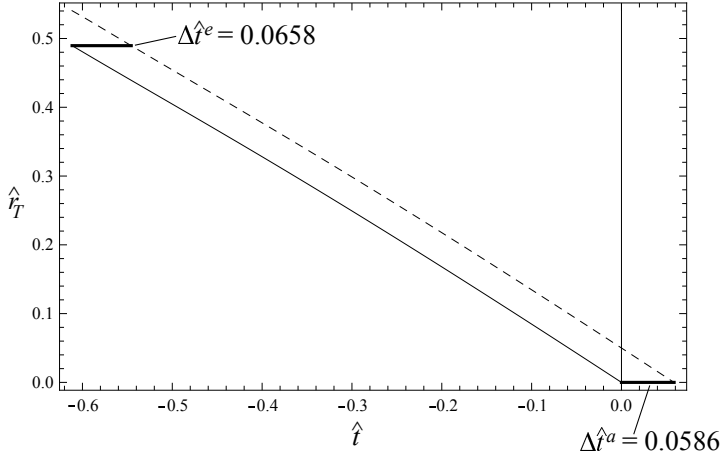}\caption{\label{fig:Two-trajectories}Two trajectories of light emitted at
$(t,r)$ (solid) and $(t+\Delta t^{e},r)$ (dashed) respectively and
absorbed at $(0,0)$ and $(\Delta t^{a},0)$. The black bars demonstrate
that the horizontal distance of the graphs has become smaller when
the light reaches the origin at $r=0$.Parameter choice A.}
\end{figure}

The Doppler redshift is not the only phenomenon that affects the frequency
of the light emitted by a star at $\left(t,r\right)$ and received
by an observer at $\left(0,0\right)$. There is also a blueshift due
to gravitation, because, after all, the light gains energy by \textquotedbl{}falling\textquotedbl{}
toward the center. In order to describe that additional frequency
shift, let us look at the trajectory $r_{T}(t)$ again, the curve
exhibited in previous figures and reproduced again in Fig. \ref{fig:Two-trajectories}.
The light -- already redshifted from $\underset{z}{\nu}$ to $\underset{x}{\nu}$
according to (\ref{eq:nue_x_over_Nue_z_and_d_Omega_x})\textsubscript{1}
-- starts at $\left(t,r\right)$ and proceeds along the lower graph
in Fig. \ref{fig:Two-trajectories} to the origin at $r=0$. Now let
there be a second trajectory of a light ray emitted at the same point
but at the later time $t+\Delta t^{e}$ and absorbed at the origin
at time $\Delta t^{a}$.

Obviously, since the trajectories are concave and roughly \textquotedbl{}parallel\textquotedbl{},
we have $\Delta t^{a}<\Delta t^{e}$. The short solid bars in Fig.
\ref{fig:Two-trajectories} illustrate and emphasize the situation.
A moment\'{ }s reflection shows that the ratio $\frac{\Delta t^{e}}{\Delta t^{a}}$
tends to the inverse of the slopes of the trajectories as $\Delta t^{e}$
tends to zero:
\begin{equation}
\frac{\Delta t^{e}}{\Delta t^{a}}\underset{\Delta t^{e}\rightarrow0}{\longrightarrow}\frac{\left.\frac{dr_{T}}{dt}\right|_{0}}{\left.\frac{dr_{T}}{dt}\right|_{t}}=\frac{\underset{x}{c}(0,0)}{\underset{x}{c}(t,r_{T}(t))}.\label{eq:Delta_t_e/Delta_t_a}
\end{equation}
Now, let the two emissions be consecutive emissions of the maxima
of a harmonic light wave so that $\Delta t^{e}=\frac{1}{\underset{x}{\nu}}$
and $\Delta t^{a}=\frac{1}{\underset{x0}{\nu}}$ hold, where $\underset{x0}{\nu}$
is the frequency of the light at $\left(0,0\right)$. In that case
we have
\begin{equation}
\frac{\underset{x0}{\nu}}{\underset{x}{\nu}}=\frac{\underset{x}{c}(0,0)}{\underset{x}{c}(t,r_{T}(t))}.\label{eq:nue_x0/nue_x}
\end{equation}

This means that the light arriving at the origin has been blueshifted,
since the right hand side of (\ref{eq:nue_x0/nue_x}) is bigger than
1, see Fig. \ref{fig:Two-trajectories}.

In a final step we ask for the frequency $\underset{z0}{\nu}$ of
the light in the Lorentz frame at the event $\left(0,0\right)$ where
the velocity is zero. In analogy to (\ref{eq:nue_x_over_Nue_z_and_d_Omega_x})\textsubscript{1}
we have
\begin{equation}
\frac{\underset{x0}{\nu}}{\underset{z0}{\nu}}=\sqrt{\left|Z(0,0)\right|},\quad\frac{\underset{x0}{\lambda}}{\underset{z0}{\lambda}}=\frac{1}{\sqrt{S(0,0)}}.\label{eq:nue_x0/nue_z0}
\end{equation}

\subsection{Summary on redshift}

\begin{figure}
\includegraphics{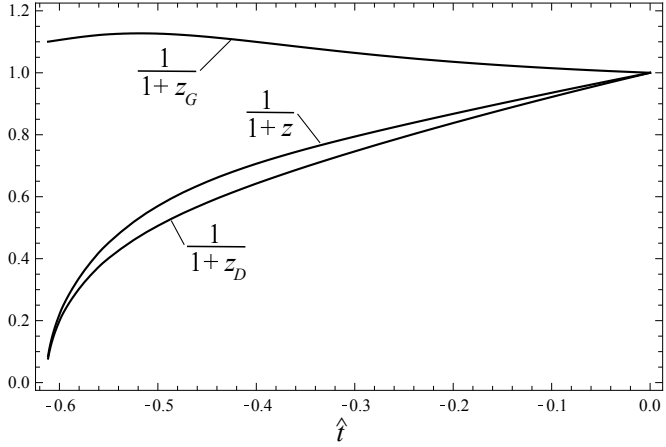}\caption{\label{fig:zG_and_zD}The contributions of Doppler shift and gravitational
shift and the overall red shift. Note that the gravitational contribution
is small compared to the Doppler contribution; the graphs refer to
choice A.}
\end{figure}

Elimination of $\underset{x0}{\nu}$ and $\underset{x}{\nu}$ between
(\ref{eq:nue_x_over_Nue_z_and_d_Omega_x}), (\ref{eq:nue_x0/nue_x}),
(\ref{eq:nue_x0/nue_z0})\textsubscript{1} provides the overall redshift
formula
\begin{equation}
\frac{\underset{z0}{\nu}}{\underset{z}{\nu}}=\underset{\frac{1}{1+z_{G}}}{\underbrace{\frac{\underset{x}{c}(0,0)}{\underset{x}{c}(t,r_{T}(0))}}}\underset{\frac{1}{1+z_{D}}}{\underbrace{\sqrt{\frac{\left|Z(t,r_{T}(t))\right|}{\left|Z(0,0)\right|}}\sqrt{\frac{1-\frac{V(t,r_{T}(t))}{\underset{x}{c}(t,r_{T}(t))}}{1+\frac{V(t,r_{T}(t))}{\underset{x}{c}(t,r_{T}(t))}}}}}.\label{eq:nue_z0/nue_z}
\end{equation}
It is customary to introduce a shift factor $z$ to replace the frequency
quotients such that $z$ is positive for a redshift and negative for
a blueshift. The definition of $z$ is indicated in (\ref{eq:nue_z0/nue_z})
both for the gravitational shift and for the Doppler shift. The overall
shift factor $z$ follows from
\begin{equation}
\frac{1}{1+z}=\frac{1}{1+z_{G}}\frac{1}{1+z_{D}}.\label{eq:1/(1+z)}
\end{equation}
Hence follows $z(t,r_{T}(t))$ for stars on the trajectory as a function
of $t$. Fig. \ref{fig:zG_and_zD} shows graphs of the two contributions
to frequency shift and the overall effect which for our choice of
parameters is a redshift.

Combining (\ref{eq:nue_z0/nue_z}), (\ref{eq:1/(1+z)}) and the definition
of $c_{x}$ from (\ref{eq:drdt_s}) we obtain
\begin{equation}
z(t)=\sqrt{\frac{S(0,0)}{S(t,r_{T}(t))}}\sqrt{\frac{1+\frac{V(t,r_{T}(t))}{\underset{x}{c}(t,r_{T}(t))}}{1-\frac{V(t,r_{T}(t))}{\underset{x}{c}(t,r_{T}(t))}}}-1.\label{eq:z_eqn}
\end{equation}
It follows that the measurement of the shift factor $z$ is not equivalent
to the measurement of the velocity of the emitting star. Indeed, since
$c_{x}$ depends on the metric components $S(t,r_{T}(t))$ and $Z(t,r_{T}(t))$,
those fields influence the relation%
\footnote{For $V=0$ equation (\ref{eq:z_eqn}) and (\ref{eq:S_from_m_r}) combine
to give the purely gravitational blueshift $z=\sqrt{1-Q\frac{\hat{m}}{\hat{r}}}-1$.
And without gravitation the equations reduce to the relativistic Doppler
shift $z=\sqrt{\frac{1+\hat{V}}{1-\hat{V}}}-1$. %
}.

\subsection{Type Ia supernovae, their luminosities and distances.}

Astrophysicists believe that Type Ia supernovae represent the collapse
of white dwarfs when they have accumulated more mass from their neighbourhood
than they can carry according to the Chandrasekhar limit. It is then
plausible to assume that these supernovae all have the same absolute
luminosity $L$, i.e. rate of energy emission. And from observations
of supernovae of known distance that absolute luminosity has the value
$L=7.5*10^{20}\:\textrm{J/s}$ corresponding to an absolute magnitude
of -19.%
\footnote{For a qualification of that categoric statement and secondary effects
that may modify it we refer the reader to the article \cite{Perlmutter2003}
by S. Perlmutter.%
} Let the emission occur at time $t$ at the distance $r_{T}(t)$ on
the trajectory of the light. We proceed to calculate the apparent
luminosity $l$, i.e. the transmission rate of energy per unit area
at $(t,r)=(0,0)$, i.e. in the center of the sphere which represents
the universe in our model.

Let $dN$ be a number of photons of energy $h\underset{z}{\nu}$ emitted
by the star at the distance $r_{T}(t)$ from the center into the element
of solid angle $d\underset{z}{\Omega}$. Because of the isotropy of
the emission in the Lorentz frame ${z^{\alpha}}$, we have $dN=\frac{E}{h\underset{z}{\nu}}\frac{d\underset{z}{\Omega}}{4\pi}$,
where $E$ is the total energy emission. The same number of photons
-- now with energy $h\underset{zo}{\nu}$ -- must pass through the
area $r^{2}d\underset{x}{\Omega}$ in the center of the sphere so
that $dN=\frac{er^{2}d\underset{x}{\Omega}}{h\underset{zo}{\nu}}$
holds, where $e$ is the energy transmission per unit area. Hence
follows
\begin{equation}
\frac{E}{h\underset{z}{\nu}}\frac{d\underset{z}{\Omega}}{4\pi}=\frac{er^{2}d\underset{x}{\Omega}}{h\underset{zo}{\nu}}.
\end{equation}
The corresponding rates of emission and transmission are
\[
L=\frac{E}{\Delta t_{z}}\:\textrm{and}\: l=\frac{e}{\Delta t_{zo}}.
\]
Elimination of $E$ and $e$ gives
\[
l=\frac{L}{4\pi r^{2}}\frac{\Delta t_{z}}{\Delta t_{zo}}\frac{\underset{zo}{\nu}}{\underset{z}{\nu}}\frac{d\underset{z}{\Omega}}{d\underset{x}{\Omega}}\:\textrm{or with}\:\frac{\Delta t_{z}}{\Delta t_{zo}}=\frac{\underset{zo}{\nu}}{\underset{z}{\nu}}
\]
\begin{equation}
l=\frac{L}{4\pi r^{2}}\left(\frac{\underset{zo}{\nu}}{\underset{z}{\nu}}\right)^{2}\frac{d\underset{z}{\Omega}}{d\underset{x}{\Omega}}.\label{eq:l_eqn}
\end{equation}
This is the desired relation between the apparent luminosity $l$
which is measurable and the absolute luminosity $L$ which is known,
see above.

We reformulate the right hand side of (\ref{eq:l_eqn}) in terms of
the redshift factor $z$:

By (\ref{eq:nue_x_over_Nue_z_and_d_Omega_x}) and (\ref{eq:nue_z0/nue_z})
through (\ref{eq:z_eqn}) we have
\[
l=\frac{L}{4\pi r_{T}(t)^{2}}\frac{S(0,0)}{S(t,r_{T}(t))}\frac{1}{(1+z)^{4}},
\]
or, in abbreviated form
\begin{equation}
l=\frac{L}{4\pi d_{L}^{2}},\:\textrm{where}\: d_{L}=r_{T}(t)\sqrt{\frac{S(t,r_{T}(t))}{S(0,0)}}(1+z)^{2}.\label{eq:d_Luminosity}
\end{equation}
$d_{L}$, defined in (\ref{eq:d_Luminosity})\textsubscript{2} will
be called the luminosity distance appropriate for our model.

In terms of apparent magnitude $m$ and absolute magnitude $M$ --
the luminosity measures preferred by astronomers -- we have%
\footnote{We have to accommodate astronomers, because they report their observations
-- like those of Fig. \ref{fig:mue-z-A} -- in a $(\mu,z)$-diagram
and they use parsec (pc) as a standard distance. We trust that the
reader will not confuse the present $m$ and $M$ with the partial
mass and the total mass which are denoted by $m$ and $M$ in the
paper elsewhere.%
}
\begin{equation}
\begin{array}{c}
\mu=m-M=5\log_{10}\left[\frac{d_{L}}{10\textrm{pc}}\right]\:\;\textrm{or}\;\:\mu(t)=5\log_{10}\left[\frac{1}{10\textrm{pc}}r_{T}(t)\sqrt{\frac{S(t,r_{T}(t))}{S(0,0)}}(1+z)^{2}\right].\end{array}\label{eq:mue_eqn}
\end{equation}

\newpage{}

\end{document}